\documentstyle[a4,12pt]{article}
\begin{document}

\def\NPB#1#2#3{{\it Nucl.~Phys.} {\bf{B#1}} (19#2) #3}
\def\PLB#1#2#3{{\it Phys.~Lett.} {\bf{B#1}} (19#2) #3}
\def\PRD#1#2#3{{\it Phys.~Rev.} {\bf{D#1}} (19#2) #3}
\def\PRL#1#2#3{{\it Phys.~Rev.~Lett.} {\bf{#1}} (19#2) #3}
\def\ZPC#1#2#3{{\it Z.~Phys.} {\bf C#1} (19#2) #3}
\def\PTP#1#2#3{{\it Prog.~Theor.~Phys.} {\bf#1}  (19#2) #3}
\def\MPLA#1#2#3{{\it Mod.~Phys.~Lett.} {\bf#1} (19#2) #3}
\def\PR#1#2#3{{\it Phys.~Rep.} {\bf#1} (19#2) #3}
\def\AP#1#2#3{{\it Ann.~Phys.} {\bf#1} (19#2) #3}
\def\RMP#1#2#3{{\it Rev.~Mod.~Phys.} {\bf#1} (19#2) #3}
\def\HPA#1#2#3{{\it Helv.~Phys.~Acta} {\bf#1} (19#2) #3}
\def\JETPL#1#2#3{{\it JETP~Lett.} {\bf#1} (19#2) #3}
\def\JHEP#1#2#3{{\it J. High Energy Phys.} {\bf#1} (19#2) #3}

\def\reflist{\section*{References\markboth
        {REFLIST}{REFLIST}}\list
        {[\arabic{enumi}]\hfill}{\settowidth\labelwidth{[999]}
        \leftmargin\labelwidth
        \advance\leftmargin\labelsep\usecounter{enumi}}}
\let\endreflist\endlist \relax

\newcommand{\be}{\begin{equation}}
\newcommand{\ee}{\end{equation}}
\newcommand{\ba}{\begin{eqnarray}}
\newcommand{\ea}{\end{eqnarray}}
\newcommand{\dal}{\raisebox{0.085cm} 
{\fbox{\rule{0cm}{0.07cm}\,}}}
\newcommand{\dslash}{{\not\!\partial}}
\catcode`\@=11
\titlepage
\begin{flushright}
LPT-ORSAY 00/81 \\
hep-th/0010179
\end{flushright}
\vskip 1cm
\begin{center}
{\Huge \bf D-branes in non-tachyonic 0B orientifolds}
\end{center}
\vskip 1cm
\begin{center}

E. Dudas  and J. Mourad
\end{center} 
\vskip 0.5cm
\begin{center}
{\it Laboratoire de Physique Th\'eorique 
\footnote{Laboratoire associ\'e au CNRS-URA-D0063.},\\
 Universit\'e de Paris-Sud, B\^at. 210, F-91405 Orsay Cedex,
 France}
\end{center}
\vskip 2cm
\begin{center}
{\large Abstract}
\end{center}
\noindent
We determine all D-branes in the non-tachyonic 0'B orientifold,
examine their world-volume anomalies and study orbifold 
compactifications. We find that the spectrum of the D-branes 
contains chiral fermions in the symmetric, antisymmetric
and fundamental representations of (unitary)  gauge groups on
the branes. The cancellation of the world-volume anomalies
requires Wess-Zumino terms which we determine explicitly.
We examine a non-tachyonic compactification to 9D whose closed 
part interpolates between 0B and IIB and revisit compactifications 
on  orbifolds. The D3-brane allows to conjecture, via the AdS/CFT
correspondence, a supergravity dual to a non-supersymmetric and 
infrared-free gauge theory. The D-string gives hints concerning the 
S-dual of the 0'B orientifold. 

\newpage
\section{Introduction}

D-branes played a major role in understanding nonperturbative 
properties
of string and field theories in the last few years. 
In addition to their
crucial role in string dualities \cite{polchinski}, 
D-branes are also at
the heart of the AdS/CFT correspondence \cite{maldacena}, 
relating gauge
field theories to string (SUGRA) theories and allowing, in
principle,
a nonperturbative formulation of supersymmetric
string theories. Some non-supersymmetric 
string theories seem to pass all the consistency 
tests which characterise the supersymmetric ones
like world-sheet superconformal and modular invariance 
and the absence of space-time anomalies. Most of these theories
have tachyons in the perturbative spectrum but there exist 
(at least) three theories which are non-supersymmtric
and with no tachyons: the $SO(16)\times SO(16)$ heterotic string,
the $USp(32)$ type I string which
is an orientifold of type IIB and  finally 
the 0'B string, an orientifold
of type 0B with a gauge group $U(32)$.
The latter has the same 
Ramond-Ramond (RR) forms as type IIB. 
So one would expect to have D-branes 
charged with respect to these
forms. On the other hand, D-branes have a conformal field theory
definition as open strings with Dirichlet boundary conditions
for a set of  coordinates \cite{polchinski}. The subjet of this
paper is to show the existence of these extended objects
in 0'B and determine their properties.
 Unlike their supersymmetric cousins,
these branes are not BPS, however the conservation of
the RR charge guarantees their stability. 

Originally conjectured to
relate superconformal field theories on D3 branes to Type IIB
string compactifications, the AdS/CFT correspondence 
was more recently 
conjectured to apply also to
non-conformal and non-supersymmetric theories coming from D3 branes 
on Type OB \cite{polyakov},
\cite{kt}, \cite{bfl2} or their orientifolds 
\cite{augusto}, \cite{carlo}, \cite{bfl}, \cite{aa}. 
The non-conformal behaviour of 
gauge theories was related to the
classical background of Type OB and to the 
closed string tachyon
condensation \cite{kt} and, for the compactifications of the
non-tachyonic orientifold
introduced in \cite{augusto}, to the dilaton 
tadpole of the model \cite{aa}.
It was argued that the gauge theory on D3 branes is free
in the ultraviolet (UV) and that the renormalization 
group running of
the gauge couplings is qualitatively in agreement 
with the dilaton (and
tachyon, whenever it exists) background on the gravity side.  

The purpose of the present paper is twofold. First, we analyze in 
detail the spectrum and the anomaly cancellation 
for all branes (D7,D5,D3 and D1) present in
the non-tachyonic OB orientifold\footnote{Properties of branes 
in Type OB were studied in \cite{bg}.}\cite{augusto}.
The consistency conditions we impose are that the world-sheet
parity operator 
squares to one on physical states and that the open string
amplitudes have a correct interpretation in the 
transverse channel as a closed string tree level exchange.
The solution to these conditions is conveniently determined 
by introducing an action of the spacetime and 
world-shhet fermions
numbers on the Chan-Paton matrices.
We find that the D-branes have a $U(D)$ gauge group
with chiral fermions in the fundamental, symmetric and
antisymmetric representations. These fermions generate
world-volume anomalies whose structure is highly 
constrained by the
ten-dimensional Green-Schwarz mechanism. The
cancellation of the anomalies allows to verify the consistency
of the world-volume theory and to predict new Wess-Zumino
couplings on the branes.

Our second purpose is to complete the study of non-tachyonic
compactifications to various dimensions. 
In particular, the simplest
non-tachyonic compactification is to 9d, where the 
massless spectrum 
of the model displays an interesting combination of Type I
compactifications with supersymmetry breaking on 
the branes (called
``brane supersymmetry breaking'' in the literature) and in the bulk 
(called ``M-theory breaking'' or ``brane supersymmetry''). 
We also study orbifold compactifications down to 8d, 6d and 4d. 
Tadpole cancellation in these compactifications 
requires the existence of D-branes, thus allowing an independent check on 
the world-volume content of the branes. 
The spectrum of D3 branes corresponds to a gauge theory which 
is actually strongly coupled in the 
UV and free in the infrared (IR).
  We  work out the 
classical background and
compare with the perturbative properties of the gauge theory. 
By working out the spectrum of the D1 brane,  
we conjecture the existence
of a corresponding (non-conventional) fundamental string theory.

The paper is organized as follows. 
We start with some notations and
conventions in Section 2. In Section 3 
we define a convenient 
formalism for finding the spectrum of non-tachyonic 
OB orientifolds and
apply it to the 10d model introduced in \cite{augusto}.
Based on this formalism we study in Section 4 
the spectrum of all D branes
and study the gauge and gravitational anomaly cancellation.
We show how the gauge part of the Wess-Zumino terms can be
obtained in a unifed manner for all D-branes.
In Section 5 we consider non-tachyonic compactifications.
First, we study the simplest example in 9D whose
closed part interpolates between 0B and IIB. Second,
we study orbifolds on $T^{2n}/Z_2$.
 We rederive the spectrum of D7,D5 and D3 
branes based on
the computation of one-loop orbifold 
amplitudes in compact space
and taking at the end the 10d limit of infinite compact volume.
In Section 6 we work out the classical background of the 
D3-D9 system
and show agreement with the massless spectrum on the D3 branes, 
via the AdS/CFT correspondence. Section 7 summarises our results and
conclusions.
 
\section{Notations and conventions}

Orientifolds \cite{cargese,ps,bs} are the appropriate context 
for a perturbative study of D
branes and their interactions. In addition to the torus
amplitude ${\cal T}$, 
at one-loop there are three additional amplitudes to
consider, the Klein bottle ${\cal K}$, 
the cylinder ${\cal A}$ and
the M\"obius ${\cal M}$. 
It is often convenient for a spacetime particle
interpretation to write the partition 
functions with the help of $SO(2n)$ characters 
\ba 
O_{2n} &=& {1 \over 2 \eta^n} 
( \theta_3^n + \theta_4^n) \ , \qquad\quad 
V_{2n}={1 \over 2 \eta^n} (
\theta_3^n - \theta_4^n) \ , 
\nonumber \\ S_{2n} &=& {1 \over 2 \eta^n} ( \theta_2^n +
i^n
\theta_1^n) \ , 
\qquad C_{2n}={1 \over 2 \eta^n} 
( \theta_2^n - i^n \theta_1^n) \ ,
\label{a1}
\ea 
where the $\theta_i$ are the four Jacobi 
theta-functions with (half)integer
characteristics. In a spacetime interpretation, 
at the lowest level 
$O_{2n}$ represents a
scalar, $V_{2n}$ represents a vector, 
while  $S_{2n}$, $C_{2n}$ represent spinors of
opposite chiralities. In order to link the direct 
and transverse
channels, one needs the transformation matrices $S$ 
and $P$ for the
level-one $SO(2n)$ characters (\ref{a1}). 
These may be simply deduced
from the corresponding transformation 
properties of the Jacobi theta
functions, and are
\be  
S_{(2n)} =  {1 \over 2}
\left(
\begin{array}{cccc}  1 & 1 & 1 & 1 
\\ 1 & 1 & -1 & -1 \\ 1 & -1 & i^{-n} & -i^{-n} \\ 1
& -1 & -i^{-n} & i^{-n}
\end{array}
\right) \ , \  P_{(2n)} = 
\left(
\begin{array}{cccc}  c & s & 0 & 0 
\\  s & -c & 0 & 0 \\ 0 & 0 & \zeta c & i \zeta s \\
0 & 0 & i
\zeta s & \zeta c
\end{array}
\right) \ ,\label{a2}
\ee   
where $c= \cos ({n \pi /4})$, $s= \sin ({n \pi /4})$ 
and $\zeta= e^{-i{n
\pi/4}}$ \cite{bs}. 
The modulus parameter $q=exp(2 i \pi \tau)$ for 
the three one-loop
surfaces is given by
\be
 {\rm Klein}: \tau = 2 i \tau_2 \ , \
 {\rm Annulus}: \tau = {i t \over 2} \ , \
 {\rm Mobius}: \tau = {i t \over 2} + {1 \over 2} 
 \ , \label{a02}
\ee
where $\tau_2 = Im \tau$ is the imaginary part of the 
closed string
torus amplitude and $t$ is the open string modulus, 
analog of the
Schwinger proper time in field theory.

In $d$ noncompact dimensions and $10-d$ compact dimensions, the relevant 
string amplitudes have the
symbolic form 
\ba
{\cal T} &\!\!=\!\!&  {1 \over (4\pi^2 \alpha')^{1+d/2}} 
\int {d^2 \tau \over \tau_2^{1+d/2}} 
{1 \over |\eta (\tau)|^{2d}} T \ , \
{\cal K}  \!=\! {1 \over (4\pi^2 \alpha')^{1+d/2}} 
\int {dt \over t^{1+d/2}}  {1 \over {\eta (\tau)}^{d}} K \ 
, \nonumber \\
{\cal A} &\!\!=\!\!& {1 \over (8 \pi^2 \alpha')^{1+d/2}}  
\int {dt \over t^{1+d/2}} {1 \over {\eta (\tau)}^{d}} A \ , \
{\cal M} \!=\! {1 \over (8 \pi^2 \alpha')^{1+d/2}}
\int {dt \over t^{1+d/2}} {1 \over {\eta (\tau)}^{d}} M \ ,
 \label{i1}
\ea
where $\alpha'$ is the string tension and 
we explicitly displayed the
contribution of the spacetime bosons.
The ``amputated'' amplitudes $T,K,A,M$ in (\ref{i1}), used for brevity 
in the rest of the paper, define the remaining part of the amplitudes,
which contain the contribution of spacetime fermions, 
of $10-d$ bosons and fermions and, in the compact space case,
the compactification lattice. We use a similar notation for Dp-branes in the
noncompact case, in which case $d$ is replaced by the dimension of the
brane world-volume, $p+1$. 

These amplitudes have the dual interpretation of 
tree-level closed
string exchanges between D branes and O planes. 
The corresponding closed
string modulus, $l$, is related to the moduli 
defined in (\ref{a02}) by
\be
 {\rm Klein}: l = {1 \over 2 \tau_2} \ , \ 
 {\rm Annulus}: l = {2 \over t}  \ , \
 {\rm Mobius}: l = {1 \over 2t} \ . 
\label{i2}
\ee 
Written in the transverse channel, 
the amplitudes (\ref{i1}) become
\ba
{\cal K} & \!\!=\!\!& {1 \over (4\pi^2 \alpha')^{1+d/2}} 
\int dl \ {1 \over {\eta (i l)}^{d}} \ {\tilde K} \ , \nonumber \\
{\cal A} &\!\!=\!\!& {1 \over (8 \pi^2 \alpha')^{1+d/2}}  
\int dl {1 \over {\eta (il)}^{d}} {\tilde A} \quad , 
{\cal M} \!=\! {1 \over (8 \pi^2 \alpha')^{1+d/2}}
\int dl {1 \over {\eta (il\!+\! 1/2)}^{d}} {\tilde M} \ , \label{i3}
\ea
where ${\tilde K}, {\tilde A}$ (${\tilde M}$) are related by the 
$S$  ($P$) transformation defined in (\ref{a2}) to 
the loop amplitudes $K,A$ ($M$). 


\section{Non-tachyonic orientifold of 0B}

The closed sector of the 0'B orientifold is given by the
0B string \cite{sw}.
The RNS world-sheet action of the 0B theory is the same as
that of the superstring with a different GSO projection, which
is given by
\be
P_{0B}={{1+(-1)^{F_{R}+F_{L}}}\over{2}},\label{gsoo}
\ee
in the Ramond-Ramond (RR)  and Neveu-Schwarz-Neveu-Schwarz (NSNS) 
sectors. Here $F_{L,R}$ are the world-sheet
fermion numbers. 
The absence of the RNS sector
leads to a space-time theory without fermions.
The presence of the $NS^-NS^-$ sector (with
$(-1)^{F_L}=(-1)^{F_R}=-1$) introduces a tachyon
in the theory. The $NS^+NS^+$ sector gives the $NSNS$
 spectrum of IIB and the $RR$ 
 sector doubles the RR spectrum of
 IIB. The spectrum and the modular invariance are 
 summarised in the torus partition function
 \be
 T=|O_8|^2+|V_8|^2+|S_8|^2+|C_8|^2.
 \ee
 
Sagnotti's $0'B$ orientifold \cite{augusto} is obtained by first gauging 
the discrete symmetry 
\be
\Omega'=\Omega(-1)^{F_{L}},
\ee
where  $\Omega$ is the usual
parity operator on the world-sheet.
Notice that
\be 
\Omega'^2=(-1)^{F_L+F_R},
\ee
which is $1$ due to the GSO projection (\ref{gsoo}).  
The tachyon is not invariant under $\Omega'$
and so is removed from the spectrum. In the $NS^+NS^+$
and the $R^+R^+$ 
 sectors
$\Omega'$ acts as $\Omega$ and so the two-form is removed
from the gravitational multiplet, 
as well as the  zero and the four-form from the first
subset of RR forms. Finally from the second subset of 
RR forms $\Omega'$ acts as $-\Omega$ and so the two-form
is removed and the axion and the  self-dual four form 
are retained.
This is summarised  in the Klein amplitude
\be
K={{1}\over{2}}(-O_8+V_8-S_8+C_8),
\ee
where the notation  was explained in Section 1.
The transverse channel Klein amplitude 
\be
\tilde K=-2^{5}  S_8,\label{kltr}
\ee
reveals a RR tadpole
for the ten-form
which has to be cancelled by the introduction of open strings.
From the spacetime point of view the  self-dual four-form leads 
to gravitational anomalies whose cancellation 
(at least their irreducible part) asks for 
chiral fermions.
The open string states carry Chan-Paton
degrees of freedom represented by matrices $\lambda$.
Two open string states may join to form a closed string state,
the amplitude being proportional to $tr(\lambda_1\lambda_2)$.
Since the closed part has no fermions, one is tempted 
to look for an open spectrum without fermions. However
this cannot be consistent because of spacetime anomalies 
from the closed spectrum. So one is forced to introduce fermions in
the open sector. The absence of fermions in the closed
sector and their presence in the open sector can be reconciled
by the decomposition of the set of Chan-Paton matrices, $\cal M$,
as a $Z_2$ graded algebra 
$\cal M=\cal M^+\oplus\cal M^-$. Recall that this means that 
an element of $\cal M$ is either an element of $\cal M^+$ 
(a boson) or an element of $\cal M^-$ (a fermion)
and that the product respects the graduation
$\cal M^+\times\cal M^+\in \cal M^+$,
$\cal M^-\times\cal M^-\in \cal M^+$ and
$\cal M^+\times\cal M^-\in \cal M^-$. 
Let the fermions have Chan-Paton
matrices in $\cal M^-$  and the bosons in $\cal M^+$.
Then $(-1)^{G}$, where $G$ is the spacetime fermion number,
 has an action on the Chan-Paton factors
\be
(-1)^{G_{cp}}(\lambda)=\sigma_9\lambda\sigma_9^{-1} \ ,
\ee
with $\sigma_9$ a unitary matrix which verifies\footnote{There is phase
choice here, that we choose to be equal to one.} $\sigma_9^2=1$.
The Chan-Paton matrices of fermions  verify
\be
(-1)^{G_{cp}}(\lambda_f)=-\lambda_f \ ,
\ee
and those of bosons
\be
(-1)^{G_{cp}}(\lambda_b)=\lambda_b \ ,
\ee
that is the action of $(-1)^{G}$ on the whole state
(oscillators, $G_o$, and Chan-Paton matrices, $G_{cp}$) gives 
1 for both fermions
and bosons. This resolves the problem of the joining of
one fermionic and one bosonic open string into a closed one.
In fact the amplitude is proportional to 
\be
tr(\lambda_f\lambda_b)
=
tr(\sigma_9\lambda_f\sigma_9^{-1}\sigma_9\lambda_b\sigma_9^{-1})
=-tr(\lambda_f\lambda_b)=0 \ .
\ee
Let the action of $\Omega'$ on the Chan-Paton matrices be given
by $\gamma_9$,
\be
\Omega' (\lambda)=\gamma_9\lambda^T\gamma_9^{-1} \ .
\ee
Then since one has $\Omega^2=1$ on the oscillators,
$\gamma_9$ has to satisfy
\be
\gamma^T_9=\epsilon\gamma_9 \ ,
\ee
where $\epsilon =\pm 1$.
Another constraint on $\gamma_9$ comes from 
$(-1)^G \Omega'=\Omega' (-1)^G$, 
\be
\sigma^T_9=\eta\gamma_9^{-1}\sigma_9\gamma_9 \ ,
\ee
where $\eta=\pm 1$.
More constraints are  obtained from the
requirement of the tadpole cancellation.
The cylinder amplitude  is given by
\be
A={{1}\over{2}} Tr \biggl\{ (-1)^{G_o}{{1+(-1)^{F}}\over{2}}
{{1+(-1)^{G_o+G_{cp}}}\over{2}}e^{-\pi tL_0} \biggr\} \ ,
\ee
where the trace is on the open sector modes. The result is
\be
A={{1}\over{4}} \biggl\{
[(tr1)^2+(tr\sigma_9)^2] V_8-
[(tr1)^2-(tr\sigma_9)^2] S_8 \biggr\} \ . \label{c}
\ee
The arguments of the characters were defined in Section 1.
Here we have used the GSO projection on the open sector 
$(-1)^F=1$, a more general setting in the next section
will clarify this choice. Here, we note that this GSO
projection is compatible with the single valuedness of the
operator product expansion as well as with the absence of non
diagonal sectors in the closed sector.

The Mobius amplitude is given by
\be
M={{1}\over{2}} Tr \biggl\{ \Omega(-1)^{G_o}{{1+(-1)^{F}}\over{2}}
{{1+(-1)^{G_o+G_{cp}}}\over{2}}e^{-\pi tL_0} \biggr\} \ , 
\ee
where the trace in on the open sector modes. We find
\be
M=-{{1}\over{4}}\epsilon [ (1+\eta)(tr 1) V_8-
(1-\eta)(tr 1) S_8 ] \ . \label{m}
\ee
From (\ref{c}) and (\ref{m}) we deduce the transverse channel
amplitudes
\ba
\tilde A={{2^{-5}}\over{4}}
[ (tr1)^2( V_8-
S_8)+(tr\sigma_9)^2( O_8- C_8)] \ , \nonumber \\
\tilde M=-{{1}\over{2}}\epsilon [
(1+\eta)(tr 1) V_8- (1-\eta)(tr 1) S_8] \ . 
\ea
The RR tadpole cancellation gives the constraints
\be
tr\sigma_9=0 \quad , \quad  
2^{-5}(tr1)^2-2\epsilon(1-\eta)(tr 1)+4\times2^{5}=0 \ .
\ee
The first constraint together with $\sigma_9^2=1$
allows $\sigma_9$ to be chosen such that $\sigma_9=
diag(1_{N\times N}, -1_{N\times N})=1\otimes\sigma_3$.
The second constraint implies that
$\epsilon=1, \eta=-1$ and $N=32$. The first two conditions can be
solved by $\gamma_9=1\otimes\sigma_1$.
This gives the gauge bosons with Chan-Paton factors
$\lambda_v=diag (h,-h^T)$, with $h$ a hermitian $32\times 32$
matrix and so the D9 gauge group is $U(32)$.
The Chan-Paton matrices of the Weyl fermions are of the form
\be
\lambda_f=\pmatrix{0 &A\cr -A^* &0} \ ,
\ee
with $A$ an antisymmetric matrix. The fermions thus belong to
 the antisymmetric representation of $U(32)$.
Note that the tadpole from the NSNS sector is non vanishing.
It induces a term in the effective action from the disk diagram
\be
\Delta S_{tadpole}=-32 T_{9}\int \sqrt{-g}e^{-\phi} \ , \label{tadpole}
\ee
where $T_9$ is the tension of the D9 branes.
The consistency of the model from the spacetime point of view
can be verified by calculating the total anomaly due to
the fermions from the open sector and the self-dual form from the
closed sector.
The anomaly polynomial\footnote{Here the actual
value of the anomaly polynomial is ${\hat X}_{12} = 2 \pi X_{12}$. We
shall equally use this convention in the rest of the paper.} has the form \cite{augusto}
\be
X_{12}=X_2X_{10}+X_4X_8+{{1}\over{2}}X_6^2 \ , \label{ans}
\ee
with
\ba 
X_2=itrF, \ X_4=-{{1}\over{2}}\left(trF^2-trR^2\right) \ , \nonumber \\
 X_6=-{{i}\over{6}}\left(trF^3-{{1}\over{8}}trF tr R^2\right) \ ,
\nonumber \\
X_{8}={{1}\over{24}}\left(trF^4+{{1}\over{8}}tr R^4
 +{{1}\over{32}} (trR^2)^2-{{1}\over{8}}tr F^2tr R^2\right) \ , \nonumber \\
 X_{10}={{i}\over{48}}
 \left({{2}\over{5}}tr F^5+{{1}\over{240}}trF tr
 R^4-{{1}\over{192}}trF(tr R^2)^2\right) \ . \label{ans1}
 \ea
The irreducible anomaly in $trR^6$ has cancelled
and this particular factorised form allows a Green-Schwarz
mechanism to cancel the remaining anomaly. The Green-Schwarz
terms in the effective action read
\be
\Delta S_{GS}=T_{-1}\int_{D9} A_0X_{10}+ T_{1}\int_{D9}A_2X_8
+{{T_3}\over{2}}\int_{D9} A_4X_6 \ , \label{GS}
\ee
 where $A_0, A_2$ and $A_4$ are respectively the RR axion,
 two-form and self dual four-form. We shall see in the next
 section that $T_i$ is the tension of a $i$-brane.
 The Green-Schwarz mechanism is
 completed by the modification of the Bianchi identity of these
 forms 
 \be
 dF_{i+1}={{2\pi}\over{T_{i-1}}} X_{i+2} \ , \ i=0,2,4 
 \ ,\label{bia}
 \ee
 so that the variation of the forms under gauge transformations
 are given by
 \be
 \delta A_i=-{{2\pi}\over{T_{i-1}}} X_i^{1}\ ,\ i=0,2,4\ \ ,
 \label{varia}
 \ee
 where the $X_{i}^{1}$ are related to $X_{i+2}$ by the descent
 equations
 \be
 X_{i+2}=dX_{i+1},\ \delta X_{i+1}=dX^{1}_{i}\ .
 \ee
 The field equations of the RR forms are modified due to
(\ref{GS}) 
 \be
 d^*F_{i+1}=2T_{i-1}\kappa^2 X_{10-i} \quad , \quad i=0,2,4 \ , \label{eqf}
 \ee
 where $\kappa^2$ is the ten-dimensional 
 gravitational constant. Notice that the factor $1/2$
 in (\ref{GS}) in front of the term in $A_4$
 is essential in obtaining the field equation (\ref{eqf})
 for the self-dual form\footnote{This is easily verified by going
 to 9 dimensions, where the self-dual form gives one three-form, or more
 covariantly by using the PST formalism \cite{pst}.}. 
 The  self-duality of $F_5$ implies that $T_3$ has a fixed 
 value
 \be 
 T_3^2={{\pi}\over{\kappa^2}} \ , \label{ten3}
 \ee
 a result which we shall derive from a string calculation
  of the amplitude between two branes in the next
 section. Note that the Dirac quantization condition (with minimum charge) 
 $T_iT_{6-i}\kappa^2=\pi$ allows to write both the field equations
 and the Bianchi identities as (\ref{bia}) with the index $i$
 taking even values from $0$ to $8$.
 We shall  show in the next section how each polynomial
 $X_{i+2}$ gives the irreducible world-volume 
 anomaly of the $p=i-1$ D-brane, providing a remarkable
 relation between space-time and world-volume anomaly
 cancellation. The basic idea is that the variations
 (\ref{varia}) induce a non-trivial transformation of the
 Wess-Zumino term of the $D$ brane
$T_{i}\int_{D} A_{i+1}$ which has to be cancelled by the
world-volume anomalies.
This is similar to the relation between 
the usual Green-Schwarz mechanism and the world-volume
anomalies of the string \cite{hw} and the five-brane
\cite{ddp,lt,mourad,lm}. 
 \section{0'B D-branes}
 
 From the RR forms of 0'B we expect to have 
 D$(2p+1)$ branes for $p=-1$,$0$,$1$,$2$,$3$,$4$. The 9-branes 
 were considered in the previous section. Here
  we determine from the world-volume point of view 
  the consistent theories  on the branes.
  There are essentially two consistency conditions:
  first $\Omega$ should square to one on the 
  Dirichlet-Dirichlet (DD) and the Neumann-Dirichlet (ND) 
  states and second the cylinder and Mobius amplitudes should
  have a correct
  interpretation in the transverse channel as an exchange of
  physical
  closed string states.
  We shall confirm the massless excitations from the 
  world-volume anomaly
  analysis in this section as well as from the explicit
  computation of the string amplitudes in orbifold
  compactifications in the next section.
  
  The excitations of the D-branes come from 
  two sectors: the DD sector where the open string has both ends
  on the D-brane and the ND sector with one end  on the brane
  and the other  on the 9-brane. As before, there are bosonic and fermionic
  Chan-Paton factors distinguished by the value of 
  $(-1)^{G_{cp}}$, which
  is $1$ in the NS sector and $-1$ in the $R$ sector.
  We denote the corresponding matrix by $\sigma_p$. 
  Similarily, to the world-sheet fermion number $(-1)^F$
  we introduce an action on the Chan-Paton factors
  \be
  (-1)^{F_{cp}}(\lambda)=f_p\lambda f_p^{-1} \ ,
  \ee
  where $f_p$ is a unitary matrix which squares to one
  and commutes  with $\sigma_p$. The presence of this action
  on the Chan-Paton factors is needed in order to insure that 
  two open strings join to form a closed string with only 
  diagonal sectors.
  
  The action of $\Omega$ on the oscillator degrees of freedom
  in the DD sector is the same as the action in the NN
  sector except that there is in addition 
  a spacetime parity operation on the transverse 
  space to the brane, that can be written symbolically as
  \be 
  \Omega_{DD}=\pi \Omega_{NN} \ .
  \ee
  When acting in the R sector, the zero mode part of the parity
  is given by $\pi_{0}=\Pi_{i} (\Gamma\Gamma^i)$, where
  $i$ labels the $9-p$ transverse coordinates and  $\Gamma$ is the
  ten dimensional chirality matrix. 
  The square of $\pi_{0}$ is given by
  \be
  \pi_{0}^2=(-1)^{(9-p)(10-p)/2},
  \ee
  which is $+1$ for $p=4n+1, 4n+2$ and $-1$
  for $p=4n, 4n+3$.  For the total action of 
  $\Omega^2$ to be one
  it is necessary that the action of $\Omega^2$ 
  on the fermionic Chan-Paton factors be given
  by the value of $\pi_{0}^2$. Therefore, if the action of $\Omega$ 
  on the DD Chan-Paton matrices is given by $\gamma_p$, we have
  \be 
  \gamma_p(\gamma_p^{-1})^{T}=\epsilon_p,\label{cons}
  \ee
  for $p=4n+1, 4n+2$ and
  \be
  \gamma_p(\gamma_p^{-1})^{T}=\epsilon'_p\sigma_p \label{constr}
  \ee
  for $p=4n, 4n+3$. Equation (\ref{cons})
  implies that $\epsilon_p=\pm 1$ and equation 
  (\ref{constr}) may be rewritten as
  \be
  \gamma_p=(\epsilon_p')^2\sigma_p \gamma_p \sigma_p^T,\label{ee}
  \ee
  which, using $\sigma_p^2=1$ gives $(\epsilon_p')^4=1$.
  The values of $\epsilon_p$ and $\epsilon'_p$ are determined 
  from the ND sector. Here also we have to 
  impose that the square
  of $\Omega$ be 1 on the tensor product of the oscillator part
  and the Chan-Paton part of the states. 
  The action of $\Omega^2$ on the oscillators can be deduced,
  following \cite{gp},
    by considering the product of
   a ND vertex operator and a DN vertex operator. 
  In the R sector, where the supercurrent $T_F$ is periodic, 
  $\Omega^2$ acts as
  $i^{(9-p)/2}$ and in the NS sector, where $T_F$ is
  antiperiodic, $\Omega^2$ acts as $i^{(p-1)/2}$.
  On the other hand, the Chan-Paton factors in the 
  Ramond ND sector satisfy
  \be 
  \sigma_9\lambda\sigma_p^{-1}=-\lambda.\label{g}
  \ee
   The Neveu-Schwarz ND Chan-Paton factors obey a similar relation
   except for a plus sign on the right hand side.
   The action of $\Omega^2$ on the Ramond ND Chan-Paton factors 
   must satisfy  
  \be
  \gamma_9(\gamma_9^{-1})^{T}\lambda\gamma_p^{T}\gamma_p^{-1}
  =(i)^{(p-9)/2}\lambda\label{om}
 \ee
 and in the NS sector they  must satisfy  
  \be
  \gamma_9(\gamma_9^{-1})^{T}\lambda\gamma_p^{T}\gamma_p^{-1}
  =(-i)^{(p-9)/2}\lambda \ . \label{omr}
 \ee
  The tadpole cancellation gave $\gamma_9(\gamma_9^{-1})^{T}=1$.
  Consider the case $p=4n+1$, then using (\ref{cons}),
  equations (\ref{om}) and (\ref{omr}) become
  \be
  \lambda=\epsilon_p(-1)^{n}\lambda \ ,
  \ee
  which gives $\epsilon_{1}=1, \epsilon_{5}=-1$ for the the D
  string and the D5-brane. Now consider the case $p=4n+3$,
  then (\ref{constr}) allows (\ref{om}) and (\ref{omr}) to 
  be written as
  \be
  \lambda \sigma_p^{-1}=
  i\epsilon'_p(-1)^{n}\sigma_9\lambda\sigma_p^{-1} \ , \label{eee}
  \ee
  where we used the action of $(-1)^{G}$ on the Chan-Paton factors.
  Multiplying (\ref{eee}) by $\sigma_p$
  leads to
  $(1-i(-1)^{n}\epsilon'_p\sigma_p)\lambda=0$. 
  This is consistent
  provided $\epsilon'_p=\pm i$. Finally we get
  $\epsilon'_3=\pm i$ and $\epsilon'_7=\pm i$.
  The commutation of $(-1)^{G}$ and $(-1)^F$ with $\Omega$ 
  gives
  \be 
  \sigma_p^T=\eta_p\gamma_p^{-1}\sigma_p\gamma_p,
  \ f_p^T=\eta_p'\gamma_p^{-1}f_p\gamma_p \ , \label{com}
  \ee
  whose compatibility with (\ref{ee}) for $p=3$ 
  and $p=7$ gives 
  \be
  \eta_p=\epsilon'_p{}^2=-1 \ , \  p=3,7 \ . \label{eta}
  \ee  
  
  So far the constraints were obtained from
  the absence of non-diagonal closed strings  and the 
  existence of states where $\Omega$ squares to one 
  on the tensor product of Chan-Paton and oscillator
  degrees of freedom. There are other constraints 
  which come from
  the requirement that the open string diagrams
  (the cylinder and the Mobius) have a dual interpretation. 
  Indeed, after a modular transformation the diagrams have
  the interpretation of tree level closed strings diagrams.
  Consider the cylinder amplitude for DD open strings
  \be
  A_{p-p}={{1}\over{2}}Tr\Big\{(-1)^{G_o}{{1+(-1)^{F_o+F_{cp}}}
  \over{2}} {{1+(-1)^{G_o+G_{cp}}}
  \over{2}} e^{-\pi tL_0}\Big\} \ ,
  \ee
  where the trace is on the DD open string states. 
  It reads
  \ba
  A_{p-p}&=&{{1}\over{8}} \biggl\{ [
  (tr1)^2+(tr\sigma_p)^2+(tr f_p)^2+(tr f_p\sigma_p)^2]
   V_8 \nonumber\\
  &+& [
  (tr1)^2+(tr\sigma_p)^2-(tr f_p)^2-(tr f_p\sigma_p)^2]
  O_8\nonumber\ \\
  &-& [
  (tr1)^2-(tr\sigma_p)^2+(tr f_p)^2-(tr f_p\sigma_p)^2]
  S_8\nonumber\\ &-& 
  [
  (tr1)^2-(tr\sigma_p)^2-(tr f_p)^2+(tr f_p\sigma_p)^2]
  C_8 \biggr\}{{1}\over{\eta^{9-p}}} \ . \label{cyli}
  \ea
  A modular transformation brings the amplitude to the form
  \be
  \tilde A_{p-p}={{2^{-(p+1) / 2}} \over{4}} l^{p-9 \over 2}
  \left[ (tr1)^2 V_8
  +(tr\sigma_p)^2  O_8 -(tr f_p)^2 S_8
  -(tr f_p\sigma_p)^2 C_8
  \right] {{1}\over{\eta^{9-p}}} \ . \label{tran}
  \ee
  The character $ O_8$ has the interpretation of the exchange between
  the branes of closed string states whose lowest mass is the
  tachyon. The latter has been removed by the $\Omega'$ projection 
  so this term in the transverse channel amplitude has to be
  absent. This is possible if
  \be
  tr\sigma_p=0 \ , \label{tr}
  \ee
  which is another consistency requirement on $\sigma_p$.
  By a suitable change of basis $\sigma_p$ may be put in the form
  $\sigma_p=1_{M\times M}\otimes \sigma_3$.
  The one and five brane couple to RR fields that originate
  from the $|S_8|^2$ sector of 0B, whereas the 3 and 7 brane
  couple
  to the RR forms that come from the 
  $|C_8|^2$ sector, so we get the constraints
  \ba
  tr(f_p\sigma_p)=0& , & \ p=1,5 \ , \label{tr2}\\
  tr(f_p)=0&,& \ p=3,7 \ . \label{tr3}
  \ea
  The Mobius amplitude also has to have a correct transverse
  channel interpretation as an exchange of closed string states
  between the $O9$ and the D-brane. 
  The direct channel amplitude
  reads
   \ba
  M_{p}&=&-{{1}\over{8}}(1+\eta_p)
  tr(\gamma_p^{-1}\gamma_p^T)
  \biggl[(1+\eta'_p)
  (V_{p-1}O_{9-p}-O_{p-1}V_{9-p})\nonumber\\
  &+&(1-\eta'_p)(O_{p-1}O_{9-p}-V_{p-1}V_{9-p})
   \biggr]{{1}\over{\eta^{9-p}}}\nonumber \\
  &+&{{1}\over{8}}i^{(9-p)(10-p)/2}(1-\eta_p)
  tr(\gamma_p^{-1}\gamma_p^T)\biggl[(1+\eta'_p)
   (S_{p-1}S_{9-p}-C_{p-1}C_{9-p})
   \nonumber\\
    &+&(1-\eta'_p)
   (C_{p-1}S_{9-p}-S_{p-1}C_{9-p})
   \biggr]{{1}\over{\eta^{9-p}}} \ ,
  \ea
  and a $P$ transformation gives the transverse amplitude 
  \ba
  \tilde M_{p}&=&-{{1}\over{4}}(1+\eta_p)
  tr(\gamma_p^{-1}\gamma_p^T)
  \biggl[(1+\eta'_p)
  \biggl(\cos({{{p-1}\over{2}} \pi})
 (V_{p-1}O_{9-p}-O_{p-1}V_{9-p})\nonumber\\
 &-&\sin({{{p-1}\over{2}}\pi}) O_8
\biggr) 
  -(1-\eta'_p) (O_{p-1}O_{9-p}-V_{p-1}V_{9-p})\biggr]
  {{1}\over{\eta^{9-p}}}
   \nonumber \\
  &+&{{1}\over{4}}(1-\eta_p)
  tr(\gamma_p^{-1}\gamma_p^T)\biggl[(1+\eta'_p)
   \biggl(\cos({{{p-1}\over{2}}\pi})
   (S_{p-1}S_{9-p}-C_{p-1}C_{9-p}) \nonumber\\
   &+&i\sin({{{p-1}\over{2}}\pi}) (C_{p-1}S_{9-p}-S_{p-1}C_{9-p})
   \biggr)\nonumber\\
   &+&(1-\eta'_p)\biggl(
   \cos({{{p-1}\over{2}}\pi}) (C_{p-1}S_{9-p}-S_{p-1}C_{9-p})
   \nonumber\\
   &+&i\sin({{{p-1}\over{2}}\pi}) (S_{p-1}S_{9-p}-C_{p-1}C_{9-p})
   \biggr) \biggr]{{1}\over{\eta^{9-p}}} \ .
   \ea
  The $O9$ plane couples only to the RR fields 
  originating from the $R^+R^+$ sector, so the Mobius amplitude
  in the transverse channel must only contain 
  the contribution of $S_8$. On the other hand the 
  D3 and D7 branes couple to $R^-R^-$ fields, so the Mobius
  amplitude must vanish for these branes and indeed this is the
  case due to (\ref{constr}) and (\ref{tr}).
  For the other branes, (the D1 and D5 branes),
  we get the additional constraints
  \be 
  1+\eta_p=0,\ 1-\eta'_p=0,\ p=1,5 \ . \label{et}
  \ee
  The first insures that the Mobius
  amplitude does not contain  an exchange of NSNS fields
  and the second that it contains no $R^-R^-$ modes.
  Notice that (\ref{eta}) and (\ref{et})
  imply that  $\eta_p=-1 \ , \forall p$.
  
  It is consistent to ask for the absence
  of open string tachyons for states stretched
  between two branes. In fact, the conditions
  \ba
  f_p&=& 1,\ p=1,5\label{nota1} \\
  f_p&=&\sigma_p \ , \ p=3,7 \ , \label{nota2}
  \ea
  insure that the amplitudes (\ref{cyli}) do not involve the open
  string tachyon. In this case the cylinder amplitudes take the
  simple form
  \ba
  A_{p-p}&=&
  {{(tr 1)^2}\over{4}}\left(V_8-S_8\right) \ , \ p=1,5 \ , \label{ano1}\\
  A_{p-p}&=&
  {{(tr 1)^2}\over{4}}\left(V_8-C_8\right) \ , \ p=3,7 \ . \label{ano2}
  \ea
     
 It remains to consider the interaction of the D-branes with the
 9-branes filling the spacetime.
 This is described by the cylinder amplitude between 
 the 9 and the
 D brane. It is given in the direct channel by
 \ba
 A_{9-p} &=& {1 \over
 4}  \biggl\{
 \left[ (tr 1_9)(tr 1_p)+tr(\sigma_9)tr(\sigma_p)
+tr(f_9)tr(f_p)+tr(\sigma_9f_9)tr(\sigma_p f_p)\right]
\nonumber\\
&&(V_{p-1}S_{9-p}+O_{p-1}C_{9-p})
\nonumber\\
&+& \left[ (tr 1_9)(tr 1_p)\!+\!tr(\sigma_9)tr(\sigma_p)
\!-\! tr(f_9)tr(f_p)\!-\!tr(\sigma_9f_9)tr(\sigma_pf_p)\right] \nonumber \\
&& (O_{p\!-1}S_{9\!-p}\!+\!V_{p\!-1}C_{9\!-p})
\nonumber\\
&-& \left[ (tr 1_9)(tr 1_p)-tr(\sigma_9)tr(\sigma_p)
+tr(f_9)tr(f_p)-tr(\sigma_9f_9)tr(\sigma_pf_p)\right]
\nonumber \\
&&(S_{p-1}O_{9-p}+C_{p-1}V_{9-p})
\nonumber\\   
 &-&
\left[ (tr 1_9)(tr 1_p)-tr(\sigma_9)tr(\sigma_p)
-tr(f_9)tr(f_p)+tr(\sigma_9f_9)tr(\sigma_pf_p)\right]
\nonumber\\
&&(C_{p-1}O_{9-p}+S_{p-1}V_{9-p}) \biggr\}  
\left({{\eta}\over{\theta_4}}\right)^{(9-p)/2} \ . \label{ano3}
\ea
Here we have used the relations $(-1)^{F_o}=
(-1)^{F_o^{(p-1)}}(-1)^{F_o^{(9-p)}}$ and 
the convention that $O_{p-1}$ (resp. $V_{p-1}$)
corresponds to $(-1)^{F_o^{(p-1)}}=-1$ (resp. 1),
while $O_{9-p}$ (resp. $V_{9-p}$) corresponds to
$(-1)^{F_o^{(9-p)}}=1$ (resp. -1). Similarily,
$S$ and $C$ correspond respectively to  
the $+1$ and $-1$ values of $(-1)^{F_o}$.
We have also used the fact that each
antiperiodic boson contributes  with $\sqrt{\eta/\theta_4}$ 
to the partition function.
 After using (\ref{tr}, \ref{tr2}, \ref{tr3})
and (\ref{nota1}, \ref{nota2}) as well as
$f_9=1$ and $tr\sigma_9=0$, (\ref{ano3}) simplifies to
\ba
 A_{9-p}&\!\!\!=\!\!\!&{ (tr 1_9)(tr 1_p) \over{2}}
 \left[ V_{p-1}S_{9-p}\!+\!
O_{p-1}C_{9-p} \!-\!S_{p-1}O_{9-p}\!-\!C_{p-1}V_{9-p} \right] \nonumber \\
&\times& \left({{\eta}\over{\theta_4}}\right)^{9-p \over 2} \ , \ 
{\rm for} \ p=1,5 \nonumber \\
A_{9-p}&\!\!=\!\!&{{(tr 1_9)(tr 1_p)}\over{4}}
\biggl\{
(V_{p-1}\!+\!O_{p-1}) (S_{9-p}\!+\!C_{9-p}) \nonumber \\
&-& (S_{p-1}\!+\!C_{p-1}) (V_{9-p}\!+\!O_{9-p}) \biggr\} 
\left({{\eta}\over{\theta_4}}\right)^{9-p \over 2} \ ,   
\ {\rm for } \ p=3,7 \ .
\ea
The transverse amplitudes read
\ba
&\tilde A_{9-p}&= {2^{-(p+1)/2} \over 2} (tr 1_9)(tr 1_p)
\biggl\{ V_{p-1} O_{9-p}\!-\! O_{p-1} V_{9-p} \nonumber \\
&-& (-1)^{(9-p)/4}( S_{p-1} S_{9-p}\!-\!C_{p-1}C_{9-p}) \biggr\} 
\left({{\eta}\over{\theta_2}}\right)^{9-p \over 2}
\ea
for $p=1,5$ and
\be
\tilde A_{9-p}={2^{-(p+1)/2} \over 4} (tr 1_9)(tr 1_p)
\left[ V_{p-1} O_{9-p}\!-\! O_{p-1} V_{9-p}\right] 
\left({{\eta}\over{\theta_2}}\right)^{9\!-\!p \over 2}
\ee
for $p=3,7$, which indeed represent the exchange of physical closed string
modes between the Dp and the D9 brane.

Notice that  (\ref{ano1},\ref{ano2}) imply that 
the cylinder amplitude between two identical branes is numerically 
zero. That is, there is a 
boson-fermion degeneracy in the open sector and a corresponding
cancellation between the NSNS  and RR closed string exchange.
On the other hand, the interaction between the 9-brane and the
D-brane, given by $\tilde A_{9-p}$  is non-vanishing
except for $p=5$.
Let $N_0(p)$ be the minimum rank of the matrices representing 
the Chan-Paton matrices. Then the tension of the branes can be
deduced from the  transverse amplitude (\ref{ano1},\ref{ano2})
to be
\be
T_{p}^{0'B} = T_{p} {{{N_0(p)}\over{{2}}}} \ , \label{ten}
\ee
where $T_{p}$ is the corresponding tension 
in type $IIB$ given by 
\be
T_{p}^2 = {{\pi}\over{\kappa^2}}(4\pi^2\alpha')^{3-p} \ .
\ee
  
\subsection{D3-brane}

 Since we have chosen $\sigma_3$ to be of the form
 \be
 \sigma_3=\pmatrix{1_{M \times M} &0\cr
 0 &-1_{M\times M}} \ ,
 \ee
 the bosonic Chan-Paton matrices are diagonal
 \be
 \lambda_b=\pmatrix{X_1 &0\cr
 0 &X_2} \ , \label{boso}
 \ee
 and the fermionic ones off-diagonal
 \be
 \lambda_f=\pmatrix{0 &Y_1\cr
 Y_2 &0} \ . \label{ferm}
 \ee
In the DD sector, $X_1$ and $X_2$ are hermitian and
$Y_1^{\dagger}=Y_2$. They are further constrained by
the invariance under $\Omega$. Recall that  
 $\gamma_3$  anticommutes with $\sigma_3$ and verifies
 $\gamma_3(\gamma_3^{-1})^{T}=i\sigma_3$
 \footnote{The other choice
 $\gamma_3(\gamma_3^{-1})^T=-i\sigma_3$ leads to the same spectrum
 with a global chirality change.}.
 A solution is given by
 $\gamma_{3}=1_{M\times
 M}\otimes\left({{\sigma_1+\sigma_2}\over{\sqrt{2}}}\right)$:
 \be
 \gamma_{3}=\pmatrix{0_{M\times M} &e^{i\pi/4} 1_{M \times M} \cr
 e^{-i\pi/4} 1_{M\times M} &0_{M\times M}}.\label{gam3}
 \ee
 The DD Chan-Paton matrices for the vectors obey
 $\lambda_v=-\gamma_3\lambda_v^{T}\gamma_3^{-1}$,
 with $\lambda_v$ of the form (\ref{boso}).
 This is solved by $\lambda_v=diag(h,-h^{T})$ where $h$ are hermitian
$M\times M$ matrices. The resulting gauge group is thus $U(M)$.
There are also  6 massless scalars representing the fluctuations 
of the brane. Their Chan-Paton matrices  verify
$\lambda_s=\gamma_3\lambda_s^{T}\gamma_3^{-1}$
and so have the form $\lambda_s=diag(h,h^T)$. They are in
the adjoint of $U(M)$. 
The GSO projection in the R sector is onto states with
$(-1)^{F}=-1$. The states are characterised by their four
dimensional chirality $\epsilon$ 
which is correlated with the six-dimensional
transverse one $\epsilon'=-\epsilon$.
The massless fermions with
four-dimensional chirality $\epsilon$
have Chan-Paton factors of the form (\ref{ferm}),
verifying $\lambda_{f\epsilon}=
i\epsilon \gamma_3\lambda_{f\epsilon}^T\gamma_3^{-1}$.
This is due to the fact that $\Omega$ acts as $i\epsilon '$ on the
oscillator part of the state. We get negative chirality fermions 
with Chan-Paton matrices of the form
\be
\lambda_{f-}=\pmatrix{0 &S\cr S^{*} &0} \ ,
\ee
with $S$ a symmetric complex matrix and for positive chirality
fermions
\be
 \lambda_{f+}=\pmatrix{0 &A\cr -A^{*} &0} \ ,
 \ee
 with $A$ an antisymmetric matrix. 
The fermionic spectrum
from the DD sector comprises thus, one fermion
in the $(2_-, 4_+, M(M+1)/2)$ and one
in the $(2_+, 4_-, M(M-1)/2)$ of 
$SO(1,3)\times SO(6)\times U(M)$.
This completes the massless spectrum of the DD excitations.
Note that it differs from the spectrum proposed in \cite{bfl}.
The ND sector gives massless states 
with Chan-Paton matrices $\lambda$
constrained by the condition $\Omega^2=1$  
\be
\lambda\sigma_3=\pm \lambda \ , 
\ee
where the plus sign is for the Ramond sector and the minus sign
for the Neveu-Schwarz sector. Since $F_{cp}= \sigma_3$ the GSO
projection is
onto $(-1)^{F_o}=1$ for the R sector and $(-1)^{F_o}=-1$ for the NS
sector. From the NS sector we get only masssive states.
The massless states from the R sector have negative
chirality and Chan-Paton factors of the form
\be
\pmatrix{0&0\cr m&0} \ ,
\ee
with $m$ a complex $32 \times M$ matrix. The corresponding states
transform in the $(\bar 32, \bar M)$ of $U(32)\times U(M)$
and are singlets under the transverse $SO(6)$ Lorentz group.
In the DN Ramond sector 
Chan-Paton factors must obey $\sigma_3\lambda=-\lambda$ and so
the GSO projection is
onto $(-1)^{F_{o}}=-1$. We get a massless positive chirality
fermion with a Chan-Paton matrix
\be
\lambda=\pmatrix{0&0\cr 0 &m} \ ,
\ee
that transform in the $(32,M)$ of $U(32)\times U(M)$.
Notice that the ND fermions have two degrees of freedom
as well as the DN fermions. Their invariant combination under 
$\Omega$ gives one physical positive chirality fermion in the 
$(32,M)$ of $U(32)\times U(M)$.

The smallest value $M=1$  gives $N_0(3)=2$ and thus the
tension of the 3-branes is given from (\ref{ten}) 
by $T_3^2=\pi/(\kappa^2)$, in accord with (\ref{ten3}).

Now we are in position to calculate the anomaly polynomial of the
D3-brane\footnote{This computation is very similar to the one performed
for the Type I D5 brane in \cite{mourad}.}. 
The potential anomalies are that of the transformations of
$U(M)\times U(32)\times SO(1,3)\times SO(6)$.
We will denote by $G$ (resp. $F$) the field strength 
 of the $U(M)$ (resp. $U(32)$) gauge bosons, by
 $R$ the induced space-time curvature on the world-volume and
 finally by $N$ the curvature of the normal group $SO(6)$. If
 $\Sigma$ denotes the world-volume curvature, then
 $tr R^2=tr\Sigma^2+tr N^2$.
 We shall use as  well the relations
 \ba
 Tr_A^S G &=& (M\pm 1) trG, \ Tr_A^S G^2=(M\pm 2)tr G^2+(tr G)^2 \ , 
\nonumber \\ 
 Tr_A^S G^3 &=& (M\pm 4)trG^3+3tr(G)tr(G^2) \ ,
 \ea
 where $Tr_A^S$ is the trace in the antisymmetric or symmetric 
 representation of $U(M)$ and $tr$ is as usual the trace in the
 fundamental representation.
 Using the standard anomaly formulas, as found for example 
 in \cite{gsw}, \cite{agg}, the anomaly of the world-volume fermions 
 of the D3-brane  reads
 \ba
I_6&=& -i\biggl[M{{1}\over{6}}\left(tr F^3-{{1}\over{8}}tr R^2
 trF\right)+{{1}\over{2}}tr F 
 \left({{M}\over{24}}tr N^2+tr G^2\right) \nonumber \\
&+& {{1}\over{2}}tr G\left(tr F^2-tr R^2\right) \biggr] \ .
\ea
 Notice that the irreducible part of the $U(M)$ anomaly
 has cancelled. Furthermore the anomaly 
  can be put in the particular form
 \be 
 I_6=M X_6+Y_2X_4+Y_4X_2 \ , \label{an3}
 \ee
 where 
 \be
 Y_2={i}tr G \ , \ 
 Y_4=-{{1}\over{2}}\left({{M}\over{24}}tr N^2+tr G^2\right) \ ,
 \ee
 and $X_i$ are the polynomials entering in the spacetime
 anomaly (\ref{ans}).
 From eqs. (\ref{varia}) we see that (\ref{an3})
  has the required form to be cancelled 
 by the following part of the world-volume action
 \be
 M T_3 \int_{D3}A_4+
 T_1 \int_{D3} A_2 Y_2+ T_{-1} \int_{D3}A_0Y_4 \ . \label{wz3}
 \ee
 The first term is the expected 
 RR coupling of the D3-brane, the two other terms 
 are predictions
 for the world-volume action. The second is a theta term for 
 the world-volume gauge bosons and the first gives
 mass to the $U(1)$ part of the $U(M)$ gauge fields.
 \subsection{D7-brane}
 
 Since $\gamma_7(\gamma_7^{-1})^T=-i\sigma_3$,
 the treatment of the D7-brane is somewhat
 analogous to the D-3 brane, in particular $\gamma_7$ 
 is given by the complex conjugate of
 equation (\ref{gam3}). The vectors are $U(M)$
 gauge bosons and there are 2 massless scalars in the adjoint of
 $U(M)$. A difference with respect to the D3-brane arises from the
 Ramond DD sector. If $\epsilon$ and $\epsilon'$ denote the
 chiralities of the fermions in eight and two dimensions, then
 the GSO projection being onto $(-)^{F_o}=-1$, we have
 $\epsilon=-\epsilon'$, but
 $\Omega$ now acts as $-i\epsilon'$, with a sign difference
 compared to the D3-brane. So the 77 fermions belong to
 $(8_+,1_-,M(M-1)/2)$ and $(8_-,1_+, M(M+1)/2)$
 of $SO(1,7)\times SO(2)\times U(M)$.
 From the ND sector the constraint $\Omega^2=1$ gives
 from (\ref{eee}) and $\epsilon_7'=-1$
 \be
 \lambda\sigma_3=\pm\lambda \ ,
 \ee
 where the minus sign is for the Neveu-Schwarz sector
 and the plus for the Ramond sector. 
 As for the D3-brane, this results in a 
 positive chirality fermion in the 
 $(32,M)$ of $U(32)\times U(M)$. The 97
 scalars are however now tachyons transforming in the $(32, {\bar M})$.
 
 The anomaly cancellation mechanism is more subtle than the case
 of the D3-brane because the contribution of the normal bundle is
 more involved. The anomaly polynomial
 is given by the ten-form part of
 \be
 I_{10}=\hat A(\Sigma)[
 e^{{{N}\over{2}}}Tr_A(e^{iG})-e^{-{{N}\over{2}}}
Tr_S(e^{iG})+tr(e^{iF})tr(e^{iG})] \ ,  
\ee
where $N$ is the $U(1)$ normal curvature. 
After some arrangements, explained in the Appendix, it reads
 \ba
I_{10}&=&M\left(X_{10} + {{NY_8}\over{2}}\right)
+\left(X_8+ {{Y_6N}\over{2}}\right)Y_2 +
\left(X_6 + {{Y_4 N}\over{2}}\right)Y_4 \nonumber \\
&+&\left(X_4 + {{Y_2N}\over{2}}\right)Y_6 
+ (X_2 + {{MN}\over{2}})Y_8 \ , \label{an7}
\ea
where we defined the functions (see the Appendix for more details)
\ba 
Y_2 &=& itr G \ , \ 
Y_4 = - {{1}\over{2}}tr G^2+ {{M}\over{48}}tr R^2 - {M \over 48} tr N^2 \ , 
\nonumber \\ Y_6 &=& -i\left({{1}\over{6}}tr G^3-{{1}\over{96}}  
trG tr R^2\right)- {i \over 48} tr G tr N^2 \ , \nonumber \\
Y_8 &=& {{1}\over{24}}\left(tr G^4+{{M}\over{480}}tr R^4
 -{{M}\over{384}}(tr R^2)^2 + {1 \over 4} tr G^2 tr N^2+ {M \over 320} 
(tr N^2)^2 \right)  \ ,
 \ea
where the trace in the fundamental representation of $SO(2)$ is $tr N^2 = - 2 N^2$. 
The residual anomaly is cancelled by the world-volume terms in
the action
 \be
MT_7 \int_{D7}A_8+T_5 \int_{D7}A_6Y_2+T_3 \int_{D7}A_4Y_4
+T_1 \int_{D7} A_2Y_6 +T_{-1} \int_{D7}A_0Y_8.\label{wz7}
\ee
These terms modify the equations of motion and the Bianchi
identities of the forms in ({\ref{bia}) and (\ref{eqf})
by $\delta$ terms localised on the brane
\be
 dF_{i+1}={{2\pi}\over{T_{i-1}}}(X_{i+2}+{{1}\over{2}} Y_i \delta ) \ ,
\label{delta1}
 \ee
where $\delta$ here is a two-form. This results, when 
 combined with the fact that the restriction of $\delta$ on
 the world-volume of the brane has a non-vanishing term \cite{bw}
 \be
 \delta \big|_{D7}= N \ , \label{delta2}
 \ee
 to a variation of (\ref{wz7}) which compensates
 exactly the anomaly (\ref{an7}). Notice that for the D3-brane
 the restriction of $\delta$ term to the world-volume vanishes,
 which explains the difference in the factorised structure
 of the anomaly.
\subsection{D5-brane}

For the D5-brane we have $\gamma_5^{T}=-\gamma_5$
and $\{\gamma_5,\sigma_5\}=0$, 
which are solved
by $\gamma_5=1_{M \times M}\otimes \sigma_2$.
The Chan-Paton matrices of the DD vector bosons 
are of the form $diag (h, -h^T)$, with 
$h$ a self-adjoint matrix. The
gauge group is thus $U(M)$ and acts with $g=diag(g_1,g_1^*)$.
The scalars representing the transverse fluctuations of the brane
belong as usual to the adjoint of $U(M)$.
The GSO projection in the Ramond DD sector is now onto
$(-1)^{F_{o}}=+1$ and therefore the transverse ($\epsilon'$) and
longitudinal ($\epsilon$) chiralities of
the fermions are equal. The action of $\Omega$ on 
the oscillator part of the fermions is given by $\epsilon'$,
so 
the Chan-Paton matrices of the fermions must obey
$\lambda_{f\epsilon}=\epsilon
\gamma_5\lambda^{T}_{f\epsilon}\gamma_5^{-1}$.
This is solved by
 \be
\lambda_{f-}=\pmatrix{0 &S\cr S^* &0} \ ,
\ee
where $S$ is a symmetric $M\times M$ matrix and
\be
\lambda_{f+}=\pmatrix{0 &A\cr -A^* &0} \ ,
\ee
where $A$ is antisymmetric. The DD fermions are thus in
 the $(4_-,2_-,M(M+1)/2)$ and the 
 $(4_+,2_+,M(M-1)/2)$ of $SO(1,5)\times SO(4)\times U(M)$.
In the ND  sector, the GSO projection is 
onto $(-1)^{F_o}=1$ states. From the Ramond sector,
we get one Weyl fermion of positive chirality
with Chan-Paton matrix
\be
\pmatrix{0 &\lambda_1\cr \lambda_2 &0} \ , 
\ee
where $\lambda_1,\lambda_2$ are complex $32 \times M$ matrices. The 
corresponding states transform
in the bifundamental representation of $U(32)\times U(M)$.
From the NS sector we get 2 complex massless scalars 
in the $(\bar 32, M)$.

The anomaly polynomial of the world-volume chiral fermions 
is given by
\be
I_8=M (X_8+{Y_4 \chi(N) \over 2}) +(X_6 + {Y_2 \chi(N) \over 2})Y_2
+ (X_4+{M \chi(N) \over 2})Y_4+ X_2 Y_6 \ , \label{an5}
\ee
where $\chi(N)$ is the Euler class of the normal bundle and, as
explained in detail in the Appendix, we have defined the functions
\ba
Y_2 &=& itrG \ , \ Y_4= {{M}\over{96}}tr R^2-{{1}\over{2}}
tr G^2 - {M \over 48} tr N^2 \ , \nonumber \\
Y_6 &=& -{{i}\over{6}}tr G^3 - {i \over 48} tr G tr N^2 \ .
\ea
Notice the analogy between (\ref{an5}) and (\ref{an7}). 
The anomalies are cancelled by the world-volume contribution
\be
MT_5 \int_{D5}A_6+T_3 \int_{D5}A_4Y_2+
T_1 \int_{D5}A_2Y_4+T_{-1} \int_{D5}A_0Y_6 \ . \label{wz5}
\ee
Similarly to the case of the D7 branes, eq. (\ref{delta1}), these terms 
modify the Bianchi identities. The modified Bianchi identities,
combined with the fact the restriction of $\delta$ on the D5 brane
world-volume gives \cite{bw}
\be
\delta |_{D5} = \chi (N) \ , \label{delta3}
\ee
leads to a variation of (\ref{wz5}) which exactly compensates the
anomaly (\ref{an5}).

\subsection{D string}

For the D string it is more convenient to work in a covariant
gauge.
The D string is caracterised by
a symmetric $\gamma_1$ which anticommutes with $\sigma_1$ :
$\gamma_1=1_{M\times M}\otimes \sigma_1$.
The two-dimensional gauge boson $G$ 
Chan-Paton matrices have the form
$\lambda=diag(h,-h^T)$, giving a $U(M)$ gauge group 
which acts with $g=diag (g_1,g_1^*)$. There are also 8 scalars in
the adjoint of $U(M)$, they represent the transverse fluctuations
of the string.
The GSO projection in the Ramond DD sector is onto 
$(-1)^{F_o}=1$, the longitudinal two-dimensional
and transverse eight dimensional chiralities are equal.
The action of $\Omega$ on the oscillator part is given by
$-\epsilon'$ and so
the Chan-Paton matrices must obey
$\lambda_{f\epsilon}=-\epsilon\gamma_1
\lambda^{T}_{f\epsilon}\gamma_1^{-1}$. 
The positive chirality
fermions are in the $8_+$ of the transverse Lorentz group and
have now Chan-Paton factors of the form
\be
\lambda_+=\pmatrix{0 &A\cr -A^* &0},
\ee
with $A$ an antisymmetric matrix, 
while negative chirality fermions
transform in the symmetric representation and belong to the $8_-$
of SO(8). In the ND sector, the GSO projection is onto
$(-1)^{F_o}=1$.
The ND sector gives  positive chirality fermions 
which are singlets
under the transverse $SO(8)$ and  
with matrices
belonging to the
$(\bar 32,\bar M)$ of $U(32)\times U(M)$.
The anomaly polynomial is given by
\be
I_4={{M}\over{2}}(tr R^2-trF^2)-trGtrF=MX_4
+X_2Y_2 \ ,
\ee
with 
\be
Y_2=itr G \ .
\ee
Notice the absence of terms dependent on the normal curvature.
The anomaly is cancelled by
\be
T_1 \int_{D1}A_2 + T_{-1} \int_{D_1}A_0Y_2 \ .
\ee

A particularily interesting case is for one D string.
Here the gauge group is $U(1)$,
the spectrum of the 
D string comprises in addition to the
vector boson, the scalars $X^i$, neutral under U(1),
negative chirality Weyl fermions $\theta^\alpha$,
with charge 2 under the $U(1)$
and finally 32 positive chirality Weyl fermions $\lambda^a$ with charge
$-1$ under the $U(1)$ and in the fundamental representation of
$U(32)$. 
The index $\alpha$ is a spinorial index of $SO(8)$.
The world sheet fermions from the DD sector look like the
Green-Schwarz fermions of the superstring except 
that they are now
complex. The fermions $\lambda^a$ look like the world-sheet 
heterotic fermions charged under the spacetime gauge group.
The tension of the D string being inversely
proportional to the string coupling constant,
it is natural to conjecture that in the strong coupling regime it
is the D string
which governs the perturbative dynamics. This is similar to
what happens in type I, where the D string is given by the $SO(32)$
heterotic string \cite{pw}.
The D string of the 0'B orientifold does not
have boson-fermion degeneracy in neither the right
nor the left sector so it cannot be described by a superconformal
world-sheet theory. The quantization of this string theory
is an open and interesting question. Let us just remark that the
operators $\theta^{\dagger}{}^{\alpha}\theta^{\beta}$,
$\lambda^{\dagger}{}^{a}\lambda^{b}\partial X^i$,
$\lambda^{a}\lambda^{b}\theta^{\alpha}$ are neutral
under the $U(1)$ two-dimensional gauge group and have the correct
spacetime quantum numbers of respectively the Ramond-Ramond
fields, the gauge bosons and the chiral fermions in the
antisymmetric representation of $U(32)$.
\subsection{The Wess-Zumino terms}

For simplicity we concentrate, in this subsection,
on the gauge anomalies. We would like to rederive in a unified
fashion the gauge Wess-Zumino couplings we found in the previous
subsection, revealing a remarkable structure of the 0'B D-branes.
 A concise way of writing all the Wess-Zumino interactions
 involving the gauge fields $G$ and the normal curvature $N$ is given by 
 the formula \cite{Li}
 \be
 \int A \ ({\hat A (N) })^{-1}  tr e^{iG} \ , \label{wz}
 \ee
 where the roof genus ${\hat A (N)}$ is defined in the Appendix
(\ref{roof}) and where $A$ is the formal sum of all the Ramond-Ramond forms,
 \be
 A=  \sum_{i=0}^{8} T_{i-1} A_i \ .
 \ee
 It is understood that in (\ref{wz}) only 
 the $p+1$ form term contributes for the Dp brane. Here, we shall give a
 direct proof of the gauge part in (\ref{wz}). Note first of all
 that the gauge anomaly in ten dimensions is given by the
 twelve-form part of $Tr_{A}e^{iF}$, where the trace is 
 in the antisymmetric representation. In terms of the trace in the
 fundamental representation we have the simple formula
 \be
 Tr_{A}e^{iF}={{1}\over{2}}\left[ (tre^{iF})^2-tr e^{2iF} \right] \ . 
 \label{trf}
 \ee
 The term in $tr F^6$ cancels in the above formula for $U(32)$
 since for $U(N)$ the first term gives $2N$ and the second 
 $-2^{6}$ times a common factor. The twelve form part of
 (\ref{trf}) can thus be written as the twelve form part of
 \be
 {{1}\over{2}}\left[tr(e^{iF}-1)\right]^2 \ ,
 \ee
 which is of the factorised form
 \be
 I_{12}={{1}\over{2}}\sum_{i=2}^{8}X_iX_{12-i} \ . \label{douze}
 \ee
 In (\ref{douze}),
 $ X_i$ is the $i$ form part of $ tr(e^{iF}-1)$,
 \be 
 X_i=\left(tr(e^{iF}-1)\right)\big|_i \ . \label{xf}
 \ee
 This reproduces exactly the gauge part of the 
 anomaly polynomials (\ref{ans}).
 Next, note that all the D-branes have the following
 fermion content: positive chirality fermions in the antisymmetric
 of $U(M)$, negative chirality fermions in the symmetric of
 $U(M)$,
 both of them with multiplicity $2^{(7-p)/2}$ and finally one
 multiplet of positive chirality fermions in the bifundamental of
 $U(M)\times U(32)$. Using the identity (\ref{trf})
 with a similar one for the trace in the symmetric
 (wich differs by the sign of the second term in (\ref{trf})),
 we get for the gauge anomaly $I_{p+3}$ 
 of the $Dp$-brane the $(p+3)$ form part of
 \be
 tr e^{iF}tr e^{iG}-2^{{p-7}\over{2}}tr e^{2iG} \ .
 \ee
 The term in $tr G^{(p+3)/2}$ cancels in the above formula
 and so we have
 \be
 I_{p+3}=tr (e^{iF}-1)tr e^{iG}\big|_{p+3} \ ,
 \ee
 which, using (\ref{xf}) can be written as
 \be
 I_{p+3}=\sum_iX_i tr e^{iG}\big|_{p+3} \ .
 \ee
 This is exactly the anomaly compensated by (\ref{wz}). It is also
possible, following the Appendix, to prove the form of the normal
curvature part $N$ of (\ref{wz}). Notice, however, that the $R$
curvature part of the Wess-Zumino terms cannot be written in a simple
compact form as the $N$ part in (\ref{wz}).  
 

\section{Non-tachyonic compactifications on orbifolds}

A generic compactification of $0'B$ leads to the 
reappearance of tachyons. Consider for instance the circle
compactification. The states  $|T,m,n>-|T,m,-n>$, with KK momentum 
$m$ and non-zero winding $n$, are invariant under 
$\Omega'$ but are tachyonic for small radius and zero KK momenta $m=0$. 
In the following we  examine some examples which are free from tachyons.
 
\subsection{A $9d$ example}

The simplest non-tachyonic $O'B$ compactification 
we have found is an
orientifold of the freely-acting orbifold of $OB$, generated 
by the element
$(-1)^{F_L} \times I$, where $I$ is the shift operation $I X_9 = 
X_9 + \pi r$ and $X_9$ is the last (tenth) coordinate, of radius $r$.
The torus amplitude reads
\ba
T &=& \sum_{m,n} \biggl\{ (|V_8|^2+|S_8|^2) Z_{2m,n}+  (|O_8|^2+|C_8|^2)
Z_{2m+1,n} \nonumber \\
&-& (V_8 {\bar S}_8 + S_8 {\bar V}_8) Z_{2m,n+ {1 \over 2}}
- (O_8 {\bar C}_8 + C_8 {\bar O}_8) Z_{2m+1,n+ {1 \over 2}} \biggr\} \
, \label{n1}
\ea 
where
\be
Z_{m,n} = {1 \over |\eta (\tau)|^2} 
\sum_{m,n} q^{{\alpha' \over 4} ({m \over r}+{n r \over \alpha'})^2}
{\bar q}^{{\alpha' \over 4} ({m \over r}- {n r \over \alpha'})^2} \ . \label{n2}
\ee
The amplitude (\ref{n1}) interpolates between type OB theory in the limit $r
\rightarrow \infty$ and type IIB theory in the $r \rightarrow 0$ limit.
In particular, for $r > \sqrt{\alpha'}$ we recover the tachyonic state of
type OB.   
It is however possible, starting from (\ref{n1}), to write down a Klein 
bottle amplitude which eliminates the closed tachyon for {\it all}
radii\footnote{In the following, $R$ denotes the
dimensionless radius in string units.}
\be 
K = {1 \over 2} \biggl\{ (V_8-S_8) \sum_m q^{(2m)^2 \over R^2}- 
(O_8-C_8) \sum_m q^{(2m+1)^2 \over R^2} \biggr\} {1 \over \eta} \ . \label{n3}
\ee
Notice that in the $R \rightarrow \infty$ limit and after an appropriate 
rescaling of $T$ and $K$, the closed part of the
model reproduces the 10d non-tachyonic $O'B$ orientifold \cite{augusto}.
On the other hand, in the $R \rightarrow 0$ limit we recover the closed
part of Type I superstring.
The model need for consistency the introduction of 32 D8 branes. This
can be easily justified by writing the transverse Klein amplitude
\be
{\tilde K} = 16 R \ \biggl( \sum_n q^{(2n+1)^2 R^2} V_8 - \sum_n q^{(2n)^2 R^2}
S_8 \biggr) {1 \over \eta} \ , \label{n4}
\ee 
which manifests a RR tadpole asking for 32 D8 branes.
In a spacetime interpretation, the model contains 16 $O8_{+}$ and
16 $O{\bar 8}_{-}$ planes. Since in the $R \rightarrow 0$ limit the
closed sector becomes supersymmetric, we can ask for a brane
configuration in which the RR tadpole is {\it locally} cancelled in this
limit, as in the M-theory compactifications with broken supersymmetry
studied in \cite{ads2}. The appropriate configuration has 16 D8 branes
(of Chan-Paton charge $N_1$) on top of the $O8_{+}$ planes and 16
D8 branes (of Chan-Paton charge $N_2$) on top of the $O{\bar 8}_{-}$ planes.
The open sector amplitudes are
\ba
A = \biggl\{ {N_1^2+N_2^2 \over 2}  \sum_m q^{m^2 \over R^2}
+ N_1N_2  \sum_m q^{(m+1/2)^2 \over R^2}  \biggr\} (V_8-S_8) {1 \over \eta} \ ,
\nonumber \\
M = \biggl\{ -{N_1-N_2 \over 2}  \sum_m (-1)^m q^{m^2 \over R^2} V_8
+ {N_1+N_2 \over 2}  \sum_m q^{m^2 \over R^2} S_8 \ \biggr\} {1 \over
\eta} \ . \label{n5}
\ea
The global {\it and local} tadpole cancellation ask for $N_1=N_2=16$ and
the resulting gauge group is $SO(16) \times USp(16)$. The massless
spectrum contains also Majorana-Weyl fermions in ${\bf (136,1)}$+
${\bf (1,136)}$. The resulting configuration contains  the system 
$16 \ O8_{+} - 16 \ D8$ with (massless) supersymmetric spectrum at one
of the orientifold fixed points, and the system $16 \ O{\bar 8}_{-}- 16 \ D8$
with nonsupersymmetric spectrum at the other orientifold fixed point. 
  The geometrical configuration is an interesting mixture
between the Type I compactification called M-theory breaking in 
\cite{ads2} and a 9d compactification of the Sugimoto model 
\cite{sugimoto} with a particular Wilson line. Indeed, in the M-theory 
model, the geometric configuration in 9d contained two copies of the system
$16 \ O8_{+}- 16 \ D8$, one per orientifold fixed point, each carrying
the gauge group $SO(16)$ and a massless supersymmetric spectrum.
On the other hand, by turning on a Wilson line, we can compactify to 9d 
the Sugimoto model and get a configuration with two copies of
the system $16 \ O{\bar 8}_{-} - 16 \ D8$, each giving rise to the gauge 
group $USp(16)$.  

This compactification has (for any radius ) no closed or open string 
tachyons and contains, as argued above, both the spectrum of
M-theory breaking (or brane supersymmetry) \cite{ads2} in one fixed
point and the spectrum of brane supersymmetry breaking \cite{ads}
in the other fixed point. It suggests also an interesting connection 
between the O'B orientifold introduced in \cite{augusto} and the 
Sugimoto model.

It is natural to conjecture that, as the model has the property of
satisfying {\it local} RR tadpole cancellation (notice the existence of
a dilaton tadpole for all radii), by using arguments of the type
\cite{hw,ads2}, this represents a non-supersymmetric compactification
of M-theory down to 9d which breaks supersymmetry softly in the bulk and 
on one set of the D8 branes. As the model has no closed or open tachyons
for all values of the radius, the configuration is stable at one-loop. 
This is to be
contrasted with the case of M-theory breaking \cite{ads2}, where closed
and open tachyons appear for critical value of the radius. 
It would be interesting to present the
explicit M-theory compactification realizing this configuration.

 
\subsection{0'B on $T^6/Z_2$}

In constructing string amplitudes for D9-D(9-2n) branes, we compactify
the non-tachyonic O'B orientifold \cite{augusto} on the $T^{2n}/Z_2$
orbifold ($n=1,2,3$), which reverses the sign of the $2n$ compact 
coordinates. The action in the RR sector is fixed by the action on
the zero modes, which is $Z_2 = \epsilon_n \prod_j (\Gamma
\Gamma_j) \otimes \Gamma_j$, where $\Gamma$ is the 10d chirality matrix and 
$\Gamma_j$ ($j=1 \cdots 2n$) are Dirac matrices. In addition,
$\epsilon_n$ are phases to be determined by consistency with the action
on the open sector. The orbifold
action on the $SO(2n)$ characters is
$Z_2 (O_{2n},V_{2n},S_{2n},C_{2n})= (O_{2n},- V_{2n}, i^{n(2n-1)}S_{2n},
-i^{n(2n-1)} C_{2n})$. 

We consider here string propagation on the $T^6/Z_2$
orientifold of Type O'B non-tachyonic model. Part of the D3 branes are 
moved together off the orbifold fixed points by some distance $a R$,
creating images interchanged by the orbifold action. This allows a simple way 
of finding the D3 spectrum in the noncompact space by taking the limit
of infinite compact volume and taking off the orbifold action.

We consider the orbifold acting on the $3$ compact complex coordinates as
$Z_2=(-1,-1,-1)$ and on the $SO(6)$ characters (\ref{a1}) as
$Z_2 (O_6,V_6,S_6,C_6)=(O_6,-V_6,-i S_6,i C_6)$. 

The corresponding orbifold torus amplitude, which is the starting
point for the orientifold amplitudes below, is
\ba
T&\!\!\!=\!\!\!& {1 \over 4} \biggl\{ \biggl( |O_{2} O_{6}\!+\!V_{2} V_{6}|^2 
\!+\! |O_{2} V_{6}\!+\!V_{2} O_{6}|^2 \!+\! |S_{2} S_{6}+C_{2}
C_{6}|^2 \!+\! |S_{2} C_{6}\!+\!C_{2} S_{6}|^2  \biggr) \nonumber \\
&\!\!\! \times \!\!\!& \Gamma^{(6,6)} \!\!+\!\!\! \biggl( |O_{2}
O_{6}\!\!- \!V_{2} V_{6}|^2 
\!\!+\! |O_{2} V_{6}\!\!-\!V_{2} O_{6}|^2 \!\!-\! |S_{2} S_{6}\!\!-\!C_{2}
C_{6}|^2 \!\!-\! 
|S_{2} C_{6}\!\!-\!C_{2} S_{6}|^2  \biggr) 
|{2 \eta \over \theta_2}|^{6} \nonumber \\
&\!\!\!+\!\!\!& 64 \biggl( (O_{2} S_{6}+V_{2} C_{6})\overline{(O_{2} C_{6}+V_{2} S_{6})} 
+ (O_{2} C_{6}+V_{2} S_{6}) \overline{(O_{2} S_{6}+V_{2} C_{6})} \nonumber \\ 
&\!\!\!+\!\!\!& (S_{2} O_{6}+C_{2} V_{6}) \overline{(C_{2} O_{6}+S_{2} V_{6})} +
(S_{2} V_{6}+C_{2} O_{6}) \overline{(C_{2} V_{6}+S_{2} O_{6})}   \biggr) 
|{\eta \over \theta_4}|^{6} \nonumber \\
&\!\!\!+\!\!\!& 64 \biggl( (O_{2} S_{6}-V_{2} C_{6}) 
\overline{(O_{2} C_{6}-V_{2} S_{6})}  + 
(O_{2} C_{6}-V_{2} S_{6}) \overline{(O_{2} S_{6}-V_{2} C_{6})} + \nonumber \\
&\!\!&\!\!\! (S_{2} O_{6}-C_{2} V_{6}) \overline{(C_{2} O_{6}-S_{2} V_{6})}  + 
(S_{2} V_{6}-C_{2} O_{6}) \overline{(C_{2} V_{6}-S_{2} O_{6})} \biggr)
|{\eta \over \theta_3}|^{6} \biggr\} \ , \label{a020}  
\ea
where $\theta_i$ and $\eta$ are the Jacobi and the Dedekind function, 
respectively and $\Gamma^{(6,6)}$ is the $6d$ compactification lattice 
given, in the simplest case of compact space described only by the radii 
$r_i$, by
\be
\Gamma^{(6,6)} = {1 \over |\eta (\tau)|^{12}} \sum_{m_i,n_i}
q^{{\alpha' \over 4} \sum_i ({m_i \over r_i}+{n_i r_i \over \alpha'})^2} 
{\bar q}^{{\alpha' \over 4} \sum_i ({m_i \over r_i}-{n_i r_i \over 
\alpha'})^2}
\ . \label{a0201} 
\ee
There is another consistent, modular invariant torus amplitude, 
corresponding to an inverted orbifold action on the untwisted RR 
fields compared to (\ref{a020}), i.e. $Z_2 |S_2S_6+C_2C_6|^2 \rightarrow
+ |S_2S_6- C_2C_6|^2$. This orbifold action is however incompatible with
the existence of an open sector. Indeed, as we will see below, the
orbifold acts as $\pm i$ on D3-D3 open fermions. Two such identical
fermions can join and form a closed RR state of the type $|S_2S_6|^2$
or $|C_2C_6|^2$, which must therefore be odd under the orbifold projection.  
 
The Klein bottle amplitude of the orbifold in 4d, obtained from the
torus amplitude (\ref{a020}) is
\ba  
K&\!\!\!=\!\!\!& - {1 \over 4} \biggl\{ [(\sum_m q^{m^2 \over
R^2})^6 + (\sum_n q^{n^2 R^2})^6] (O_2-V_2)(O_6-V_6) \nonumber \\
&+& [(\sum_m q^{m^2 \over R^2})^6 - (\sum_n q^{n^2
R^2})^6] (S_2-C_2)(S_6-C_6) \biggr\} {1 \over \eta^6} \ . \label{a3} 
\ea
In (\ref{a3}) and in what follows, we denote
by $m/R$ ($n R$) Kaluza-Klein (winding) masses. For simplicity, we
consider a compact space of common dimensionless radius $R^2 =
r^2/\alpha'$, where $r$ is the dimensionful radius and $\alpha'$ is the
string tension. Notice in (\ref{a3}) the absence of the twisted sector,
equivalent in the transverse channel to the absence of O9-O3 orientifold
amplitude. The amplitude (\ref{a3}) has RR tadpole divergences in the
transverse channel which ask for consistency an open sector containing
D9 and D3 branes.

We consider in the following the cylinder amplitude with D9 branes of Chan-Paton
(complex) charge $N$, part
of the D3 branes, of charge $D$, at one orbifold fixed point 
(the origin of the compact space, for simplicity) and the other D3
branes, of charge $\delta$, moved together in the bulk a distance $a R$ away
from the origin in pairs of orbifold images. The resulting amplitude reads
\ba
A &\!\!\!=\!\!\!& {1 \over 2} \biggl\{ [ N {\bar N} 
(\sum_m q^{m^2 \over R^2})^6 +
\sum_n ( (D {\bar D} + 2 \delta {\bar \delta}) q^{n^2 R^2} + 
 2 \delta {\bar \delta}  (q^{(n+2 a)^2 R^2}+ q^{(n-2 a)^2 R^2}) 
 \nonumber \\
&\!\!\!+\!\!\!& (D{\bar \delta} +{\bar D} \delta )  
(q^{(n+a)^2 R^2}+ q^{(n-a)^2 R^2}) ) (\sum_n q^{n^2 R^2})^5 ]
(O_2V_6+V_2O_6) {1 \over \eta^6} \nonumber \\
&\!\!\!-\!\!\!& {{N^2\!+\!{\bar N}}^2 \over 2 \eta^6} 
(\sum_m q^{m^2 \over R^2})^6 (S_2S_6\!+\!C_2C_6) \!-\! {1 \over 2} \sum_n
[ (D^2\!+\!{\bar D}^2 + 2 (\delta^2 + {\bar \delta}^2)) 
q^{n^2 R^2}+ \nonumber \\
&\!\!\!\!& 2 (D {\delta} \!+\! {\bar D} {\bar \delta}) (q^{(n+a)^2
R^2}\!+\!q^{(n-a)^2 R^2}) \!+\!
(\delta^2 \!+\! {\bar \delta}^2) (q^{(n+2 a)^2
R^2}\!+\!q^{(n-2 a)^2 R^2}) ] \nonumber \\
&\!\!\! \times \!\!\!& (\sum_n q^{n^2 R^2})^5 
(S_2C_6\!+\!C_2S_6) {1 \over \eta^6}\!+\! [ N
({\bar D}\!+\! 2 {\bar \delta}) (V_2S_6\!+\!O_2C_6) 
+ {\bar N} (D\!+\!2 {\delta}) \times \nonumber \\ 
&&\!\!\! (V_2C_6\!+\!O_2S_6)\!-\! N (D\!+\!2 {\delta}) (C_2V_6\!+\!S_2O_6) 
\!-\! {\bar N} ({\bar D}\!+\!2 {\bar \delta}) 
(S_2V_6\!+\!C_2O_6) ] ({\eta \over \theta_4})^3 \nonumber \\ 
&\!\!\!+\!\!\!& (Tr \gamma_N Tr \gamma_{\bar N}+Tr \gamma_D Tr
\gamma_{\bar D}) (V_2O_6-O_2V_6) ({2 \eta \over \theta_2})^3 \!+\! 
{i \over 2} \times \nonumber \\
&\!\!&\!\!\! \left[ ((Tr \gamma_N)^2\!+\! (Tr \gamma_{\bar N})^2)
(S_2S_6-C_2C_6)-  ((Tr \gamma_D)^2\!+\! (Tr \gamma_{\bar D})^2)
(S_2C_6-C_2S_6) \right] \nonumber \\
&\!\!\! \times &\!\! ({2 \eta \over \theta_2})^3
\!+\! [ -i Tr \gamma_N Tr \gamma_{\bar D} (V_2S_6-O_2C_6) +
i Tr \gamma_{\bar N} Tr \gamma_{D} (V_2C_6-O_2S_6) + \nonumber \\
&&\!\!\!\!\! Tr \gamma_N Tr \gamma_{D} (C_2V_6-S_2O_6) +Tr \gamma_{\bar N} Tr
\gamma_{\bar D} (S_2V_6-C_2O_6) ]  ({\eta \over \theta_3})^3 \biggr\} \
, \label{a4} 
\ea  
where $\gamma_N$ ($\gamma_D$)denotes the $Z_2$ orbifold action on 
D9 brane (D3 at the orbifold fixed point) Chan-Paton matrices. Notice the
peculiar structure of cylinder amplitudes, in particular the internal
chirality of fermions in the D9-D9 and D3-D3 sector are opposite,
whereas in the D9-D3 sector complex conjugation in the Chan-Paton sector exchanges
conjugate characters $S_{2,6} \leftrightarrow C_{2,6}$.  

The M\"obius amplitude reads
\ba
M &\!\!\!=\!\!\!& {1 \over 4} \biggl\{ (N\!+{\bar N}) (\sum_m q^{m^2 
\over R^2})^6 (S_2S_6+C_2C_6) {1 \over \eta^6} + \sum_n 
[(D\!+{\bar D}) q^{n^2 R^2} + \nonumber \\
&& (\delta \!+{\bar \delta}) (q^{(n+2 a)^2
R^2}+q^{(n-2 a)^2 R^2})] (\sum_n q^{n^2 R^2})^5 (S_2C_6+C_2S_6) {1 \over
\eta^6} \nonumber \\
&\!\!+\!\!& \left[ (N\!-\!{\bar N}) (S_2S_6\!-\!C_2C_6)
\!+\!(D\!-\!{\bar D}\!+\!2 \delta \!-\! 2 {\bar \delta}) (S_2C_6\!-\!C_2S_6) \right] 
({2 \eta \over \theta_2})^3 \biggl\} \ . \label{a5}
\ea
 We checked that the amplitudes (\ref{a3})-(\ref{a5}) satisfy the rules 
for orientifold construction: RR tadpole cancellation, 
factorization of amplitudes in the transverse channel and spacetime
particle interpretation.
 In the compact version of the model,  RR tadpole cancellation asks for 
the conditions
\ba
&&{\rm untwisted}: N+{\bar N} = 64 \ , \ D+{\bar D} + 2 
(\delta +{\bar \delta})= 64 \ , \nonumber \\
&& {\rm twisted}: Tr \ \gamma_N=0 \ , Tr \ \gamma_D=0 \ . \label{a05}
\ea
This, together with the appropriate particle interpretation, fixes the 
parametrization
\be
N = n_1+n_2 \ , D=d_1+d_2 \ , \label{a050}
\ee
where $n_{1,2}$ and $d_{1,2}$ are complex charges. The RR tadpole
conditions (\ref{a05}) give $n_1=n_2=16$ and $d_1=d_2=d$ and therefore
the orbifold action on the D9 and D3 (at the
orbifold fixed point) Chan-Paton charges is 
\be
\gamma_N=e^{i \pi \over 4} \ (I_{16},-I_{16}) \ ,
\gamma_D=e^{-i \pi \over 4} \ (I_{d},-I_{d}) \ , \label{a6}
\ee
where $I_{n}$ is the $n \times n$ identity matrix. The $exp(i \pi / 4)$
phase in the orbifold action is needed in order for the complete
orbifold action on the open string fermions to be a $Z_2$
action\footnote{This was already noticed in \cite{bk}, where the
spectrum of $T^6/Z_2$ OB orientifold was worked out in the compact space
and with all D3 branes at the orbifold fixed point.}. Indeed, the
orbifold action on the 99 fermion zero modes is $\pm i$
and the phase in (\ref{a6}) gives an additional action of $i$ on the
Chan-Paton factors.

The resulting gauge group is $[U(16) \times
U(16)]_9 \times [[U(d) \times
U(d)]_3 \times U(\delta)_3$, where $d+{\delta}=16$. 
Notice the usual rank reduction of the D3
gauge group due to the orbifold, which exchanges D3 branes with their
orbifold mirrors.
The massless spectrum in four dimensions consists of 
\ba
&&{\rm 6 \ scalars }: ({\bf 16},{\overline {\bf 16}};{\bf 1},{\bf 1},{\bf 1})+ 
({\overline {\bf 16}},{\bf 16};{\bf 1},{\bf 1},{\bf 1}) \nonumber \\
&+&({\bf 1},{\bf 1};{\bf d},{\bf \bar d},{\bf 1})+
({\bf 1},{\bf 1};{\bf \bar d},{\bf d},{\bf 1})+ 
({\bf 1},{\bf 1};{\bf 1},{\bf 1},{\bf{\delta {\bar \delta}}}) \ ,
\nonumber \\
&&{\rm 4 \ Weyl \ fermions }: ({\bf 120},{\bf 1};{\bf 1},{\bf 1},{\bf 1})+    
({\bf 1},{\bf 120};{\bf 1},{\bf 1},{\bf 1})+ 
({\bf 1},{\bf 1};{\bf {d(d-1) \over 2}},{\bf 1},{\bf 1}) \nonumber \\
&+&({\bf 1},{\bf 1};{\bf 1}, {\bf {d(d-1) \over 2}},{\bf 1}) +   
({\bf 1},{\bf 1};{\bf 1},{\bf 1}, {\bf {\delta( \delta-1) \over 2}})
+({\bf 1},{\bf 1};{\bf 1},{\bf 1}, \overline{{\bf {{ \delta} 
( { \delta}+1) \over 2}}}) \nonumber \\ 
&+&({\overline {\bf 16}},{\overline {\bf 16}};{\bf 1},{\bf 1},{\bf 1})+ 
 ({\bf 1},{\bf 1};{\bf \bar d},{\bf \bar d},{\bf 1}) \ , \nonumber \\
&&{\rm Weyl \ fermions }: ({\bf 16},{\bf 1};{\bf d},{\bf 1},{\bf 1})+
({\bf 1},{\bf 16};{\bf 1},{\bf d},{\bf 1}) \nonumber \\
&+& ({\bf 16},{\bf 1};{\bf 1},{\bf 1},{\bf{\delta}})+
 ({\bf 1},{\bf 16};{\bf 1},{\bf 1},{\bf{\delta}}) \ . \label{a06}
\ea

The effective field theory interactions between D-branes/O-planes and
the massless closed fields in
this model it is easily found by looking at the transverse string
amplitudes corresponding to the tree-level propagation of massless 
closed states. By using (\ref{a3})-(\ref{a5}) and denoting by ${\tilde
K}_0$, ${\tilde A}_0$,  ${\tilde M}_0$ the corresponding amplitudes,
we find, omitting unphysical couplings (for ex. proportional to 
$N-{\bar N}$), the result
\ba
&&\!\!\!\!{\tilde K}_0 \!+\! {\tilde A}_0 \!+\! {\tilde M}_0 \!=\! 
{2^{-5} \over 8} \biggl\{ [(N+{\bar N}) \sqrt{v} \!-\!{D+{\bar D}+2
\delta \!+\! 2 {\bar \delta}
\over \sqrt{v}}]^2 (O_2V_6)_0 +  \nonumber \\
&&\!\!\! [(N\!+\!{\bar N}) \sqrt{v} \!+\! {D+{\bar D}+2 \delta 
\!+\! 2 {\bar \delta} \over \sqrt{v}}]^2 (V_2O_6)_0 \nonumber \\
&\!\!\!-\!\!\!& (N\!+\!{\bar N}\!-\!64)^2 v (S_2S_6\!+\!C_2C_6)_0 \!-\!
 {(D\!+\!{\bar D}\!+\!2 \delta \!+\! 2 {\bar \delta} \!-\!64)^2 \over v}
(S_2C_6\!+\!C_2S_6)_0 \biggr\} \ , \label{a060}
\ea
where the subscript $0$ on the various characters retains only their massless
part. The corresponding effective interactions read therefore
\ba
&&\!\!\!S_{\rm int} \!=\! -{1 \over 2} T_9 \int d^{10} x \{ \sqrt{-g} 
(N\!+\!{\bar N}) e^{-\Phi} \!+\! (N+{\bar N}-64) A_{10} \} \nonumber \\
&\!\!\!-\!\!\!& {1 \over 2} T_3 \int d^4 x \{ \sqrt{-g} [D\!+\!{\bar D}
\!+\!2(\delta\!+\!
{\bar \delta})] e^{-\Phi} \!+\! [D\!+\!{\bar D}\!+\!2(\delta \!+\!
{\bar \delta})\!-\!64] A_4 \} \ , \label{a0600} 
\ea
where $A_{10}$ ($A_4$) is the ten-form (four-form) coupling to D9 (D3)
branes. The terms containing Chan-Paton factors
$N,D$ (numerical factors with $64$) describe the interactions of D9 and
D3 branes (O9 and O3 planes) with closed fields. Notice, in analogy
with the 10d counterpart (\ref{tadpole}), the unavoidable presence 
of the dilaton tadpoles, present generically in most orientifold
models with broken supersymmetry \cite{sugimoto}, \cite{ads}. 
They ask for a background redefinition \cite{fs}, explicitly performed
recently \cite{dm} for the 10d model introduced in \cite{sugimoto}.    

In the case of D3 branes in noncompact space, RR tadpoles give no
constraints on the rank of the D3 gauge group. The total gauge group
becomes $U(32)_9 \times U(M)_3$. The matter spectrum of the noncompact 
model in 4d includes the massless particles 
\ba
&&{\rm 6 \ scalars }: ({\bf 1024};{\bf 1})+ ({\bf 1};{\bf M {\bar M}}) ,
\nonumber \\
&&{\rm 4 \ Weyl \ fermions }: ({\bf 496};{\bf 1})+    
 ({\overline {\bf 496}};{\bf 1})+({\bf 1};{\bf {M(M-1) \over 2}})+  
 ({\bf 1};{\overline {\bf M(M+1) \over 2}}) \ , \nonumber \\
&&{\rm Weyl \ fermions }: ({\bf 32};{\bf M}) \ . \label{a061}
\ea
Notice that this spectrum agrees with the one obtained in the Section
4.1 by using a realization of the orientifold projection on the
Chan-Paton factors by matrices.  
It is also interesting to notice the bose-fermi degeneracy in the 33
massless spectrum.
  
\subsection{0'B on $T^4/Z_2$}

We consider here the $T^4/Z_2$ OB compact orbifold acting on the two
complex compact coordinates as $Z_2 (z_1,z_2)=(-z_1,-z_2)$ and on the
$SO(4)$ characters as $Z_2 (O_4,V_4,S_4,C_4)= (O_4,-V_4,-S_4,C_4)$. 
The open spectrum contains D9
and D5 branes. The case with D5 branes at the orbifold fixed points
was considered in \cite{carlo}. We consider here the straightforward 
extension of taking {\it all} D5 branes off the fixed points which 
encodes the spectrum of D5 branes in the noncompact space.

The corresponding orbifold torus amplitude of the model reads
\ba
T &\!\!\!=\!\!\!& {1 \over 4} \biggl\{ \biggl( |O_{4} O_{4}+V_{4} V_{4}|^2 
+ |O_{4} V_{4}+V_{4} O_{4}|^2+ |S_{4} S_{4}+C_{4}
C_{4}|^2 + \nonumber \\
&&|S_{4} C_{4}+C_{4} S_{4}|^2  \biggr) \Gamma^{(4,4)} + 
\biggl( |O_{4} O_{4}-V_{4} V_{4}|^2 
+ |O_{4} V_{4}-V_{4} O_{4}|^2+ \nonumber \\
&&|S_{4} S_{4}-C_{4} C_{4}|^2 + |S_{4} C_{4}-C_{4}
S_{4}|^2  \biggr) |{2 \eta \over \theta_2}|^{4} 
+  16 \biggl( |O_{4} S_{4}+V_{4} C_{4}|^2 \nonumber \\
&+& |O_{4} C_{4}+V_{4} S_{4}|^2+ |S_{4} O_{4}+C_{4} V_{4}|^2 +
|S_{4} V_{4}+C_{4} O_{4}|^2  \biggr) 
|{\eta \over \theta_4}|^{4}+ \nonumber \\
&16& \biggl( |O_{4} S_{4}-V_{4} C_{4}|^2 + 
|O_{4} C_{4}-V_{4} S_{4}|^2+ \nonumber \\
&&|S_{4} O_{4}-C_{4} V_{4}|^2 + 
|S_{4} V_{4}-C_{4} O_{4}|^2  \biggr)
|{\eta \over \theta_3}|^{4} \biggr\} \ , \label{a07}  
\ea
where $\Gamma^{(4,4)}$ is the $4d$ compactification lattice. 
 
The Klein amplitude in 6d is 
\ba  
K&\!\!\!=\!\!\!& - {1 \over 4} \biggl\{ [(\sum_m q^{m^2 \over
R^2})^4\!+\!(\sum_n q^{n^2
R^2})^4] [(O_4\!-\!V_4)(O_4\!-\!V_4)\!+\!(S_4\!-\!C_4)(S_4\!-\!C_4)] {1
\over \eta^4} \nonumber \\
&\!\!\!+\!\!\!& 32 [(O_4-V_4)(S_4-C_4)+ (S_4-C_4)(O_4-V_4)] 
({\eta \over \theta_4})^2 \biggr\} \ . \label{a7} 
\ea
The cylinder amplitudes with all D5 branes moved together off the fixed
points is
\ba
A &\!\!\!=\!\!\!& {1 \over 2} \biggl\{ [N {\bar N} 
(\sum_m q^{m^2 \over R^2})^4 \!+\!
2 D {\bar D} \sum_n (q^{n^2 R^2} + q^{(n+2 a)^2 R^2}+  
q^{(n-2 a)^2 R^2}) (\sum_n q^{n^2 R^2})^3 ] 
\nonumber \\
&\!\!\!\!\times\!\!\!\!& (O_4V_4+V_4O_4){1 \over \eta^4}- 
{1 \over 2} [(N^2\!+\!{\bar N}^2) (\sum_m q^{m^2 \over R^2})^4 \!+\!
(D^2\!+\!{\bar D}^2) \sum_n (2 q^{n^2 R^2}\!+\!q^{(n+2 a)^2
R^2}\!\! \nonumber \\
&\!\!+\!\!&\!\!\!\!q^{(n-2 a)^2 R^2}) (\sum_n q^{n^2 R^2})^3] 
(S_4S_4\!+\!C_4C_4){1 \over \eta^4} + 2 [ (N {\bar D} + D {\bar N}) (V_4S_4+O_4C_4) 
\nonumber \\ -
&&\!\!\!\! (N D +{\bar N} {\bar D}) (C_4V_4+S_4O_4)] 
({\eta \over \theta_4})^2  
\!+\! Tr \gamma_N Tr \gamma_{\bar N}
(V_4O_4-O_4V_4) ({2 \eta \over \theta_2})^2 \nonumber \\
&\!\!\!+\!\!\!&\! {1 \over 2} [(Tr \gamma_N)^2+ (Tr \gamma_{\bar N})^2]
(S_4S_4-C_4C_4) ({2 \eta \over \theta_2})^2 \biggr\} \ , \label{a8} 
\ea  
where here $N$ ($D$) denotes D9 (D5) brane Chan-Paton charge. 
The associated M\"obius amplitude is
\ba
M &\!\!\!=\!\!\!& {1 \over 4} \biggl\{ [(N\!+{\bar N}) (\sum_m q^{m^2 
\over R^2})^4 + (D\!+{\bar D}) \sum_n (q^{(n+2 a)^2
R^2}+q^{(n-2 a)^2 R^2}) (\sum_n q^{n^2 R^2})^3]
\nonumber \\
&&\times (S_4S_4+C_4C_4){1 \over \eta^4} +(N+{\bar N}+2D+2{\bar D})(S_4S_4-C_4C_4) 
({2 \eta \over \theta_2})^2 \biggl\} \ . \label{a9}
\ea
The RR tadpoles of the model ask for $N+{\bar N}=64$, $D+{\bar D}=32$ and
the twisted tadpoles for $\gamma_N= (i I_{16},-i I_{16})$.
The spectrum of the model in the compact space with D5 branes off the
fixed points is a straightforward generalization of the one worked out
in \cite{carlo}, in particular the gauge group is $[U(16) \times
U(16)]_9 \times U(16)_5$. 

In the noncompact version, the gauge group becomes $U(32)_9 \times U(M)_5$.
The massless charged matter content in six dimensions is
\ba
&&{\rm 4 \ scalars }: ({\bf 1024};{\bf 1})+ ({\bf 1};{\bf M {\bar M}}) ,
\nonumber \\
&&{\rm Weyl_{+} \ fermions }: ({\bf 496};{\bf 1})+    
 ({\overline {\bf 496}};{\bf 1})+ 2 \times ({\bf 1};{\bf {M(M-1) \over 2}}) \ , \nonumber \\  
&&{\rm Weyl_{-} \ fermions }: 
({\bf 496};{\bf 1})+ ({\overline {\bf 496}};{\bf 1})+ 2 \times
({\bf 1};{\bf M(M+1) \over 2}) \ , \nonumber \\
&& {\rm 2 \ complex \ scalars }: ({\bf 32};{\bf \bar M}) \ , \nonumber \\ 
&&{\rm Weyl_{+} \ fermions }: ({\bf 32};{\bf M}) \ . \label{a10}
\ea
This spectrum agrees with the one previously found in Section 4.3.  
Notice again the fermi-bose degeneracy in the 55 massless spectrum.
\subsection{0'B on $T^2/Z_2$}

The D9-D7 brane spectrum is found by starting with the compact $T^2/Z_2$
orbifold containing D9 and D7 branes and then extending the results to
the noncompact space. The analysis is very similar to the $T^6/Z_2$
orbifold considered in the D9-D3 case and will not repeated here.
The gauge group in the compact space case, with all D7 branes moved
together off the orbifold fixed points is $[U(16) \times U(16)]_9
\times U(16)_7$. 

In the case of D7 branes in noncompact space, RR tadpoles give no
constraints on the rank of the D7 gauge group. The total gauge group
becomes $U(32)_9 \times U(M)_7$. The massless spectrum of the noncompact 
model in 8d includes 
\ba
&&{\rm 2 \ scalars }: ({\bf 1024};{\bf 1})+ ({\bf 1};{\bf M {\bar M}}) ,
\nonumber \\
&&{\rm \ Weyl \ fermions }: ({\bf 496};{\bf 1})+    
 ({\overline {\bf 496}};{\bf 1})+({\bf 1};{\bf {M(M-1) \over 2}})+  
 ({\bf 1};{\overline {\bf M(M+1) \over 2}}) \ , \nonumber \\
&&{\rm Weyl \ fermions }: ({\bf 32};{\bf M}) \ . \label{a11}
\ea
In this case, however, there are tachyonic scalars in the 97 sector,
in the representations $({\bf 32};{\bf \bar M})+ 
({\overline {\bf 32}};{\bf M})$. Their presence suggests a tachyon
condensation that results, for
$M < 32$, in the unbroken gauge group $U(32-M)_9$, while for $M > 32$ the
unbroken gauge group is $U(M-32)_7$.
This spectrum agrees with the one found in Section 4.2 by using a
realization of orientifold projection on Chan-Paton factors by matrices. 
As in the previous cases, there is a bose-fermi degeneracy in the 
77 massless spectrum.
\section{Classical background of D9-D3 brane system and AdS/CFT
correspondence}

The effective action of the system in noncompact space and $D3$
gauge group $U(M)$ is easily extracted from (\ref{a0600}), by ignoring
the O3 plane contribution:
\be
S_{\rm int} \!=\! - 32 T_9 \int d^{10} x  \sqrt{-g} \ 
e^{-\Phi} - M T_3 \int d^4 x \{ \sqrt{-g} \ e^{-\Phi}+A_4 \} \ . 
\label{b1} 
\ee
In order to derive the classical background, we follow \cite{kt} which
parametrize the metric (in the Einstein frame) as  
\be
ds^2 = e^{{1\over 2} \xi (\rho)-5 \eta (\rho)} d\rho^2 + 
e^{-{1\over 2} \xi (\rho)} \eta_{\mu
\nu} dx^{\mu} dx^{\nu} + e^{{1\over 2} \xi (\rho)- \eta (\rho)} 
d \Omega_5^2 \ , \nonumber \\
\Phi = \Phi (\rho) \ . \label{b2}  
\ee

As explained in \cite{kt}, the field equations can be found from the
analog one-dimensional Lagrangian system
\be
S = \int d \rho \biggl[ {1 \over 2} ({\partial \Phi \over \partial
\rho})^2 +{1 \over 2} ({\partial \xi \over \partial
\rho})^2 - 5 ({\partial \eta \over \partial \rho})^2 - V \biggr]
\ , \label{b3}
\ee
supplemented by the constraint 
\be
 {1 \over 2} ({\partial \Phi \over \partial \rho})^2+
 {1 \over 2} ({\partial \xi \over \partial \rho})^2-
5 ({\partial \eta \over \partial \rho})^2 + V = 0 \ . 
\ee
In our case (\ref{b1}), the potential $V$ is
\be
V = - \alpha \ e^{{3 \Phi \over 2} + {1 \over 2} \xi-5 \eta}
+ 20 \ e^{-4 \eta} - M^2 e^{-2 \xi} \ , \label{b4}
\ee
where $M$ is the number of D3 branes in noncompact space and the first term
in $V$ comes from the 10d dilaton tadpole in (\ref{b1}), with $\alpha =
32 T_9 \alpha'^{5}$. The field equations derived from (\ref{b3}) read
\ba
&& \Phi^{''} = {3 \alpha \over 2} e^{{3 \Phi \over 2} + {1 \over 2} \xi-5 \eta}
 \ , \ \xi^{''} = {\alpha \over 2} e^{{3 \Phi \over 2} + 
{1 \over 2} \xi-5 \eta} - 2 M^2  e^{-2 \xi} \ , \nonumber \\
&& \eta^{''} = {\alpha \over 2} e^{{3 \Phi \over 2} + 
{1 \over 2} \xi-5 \eta} - 8  e^{-4 \eta} \ , \label{b04}
\ea
where $\Phi^{''} \equiv \partial^2 \Phi / \partial \rho^2$, etc.  There
is no known analytical solution for the coupled system of
eqs. (\ref{b04}), however approximate solutions can be found in the IR
($\rho \rightarrow \infty$) and in the UV ($\rho \rightarrow 0$). 
Notice first of all that, from (\ref{b04}) the dilaton increases
at larger distances from the brane. This suggests that the gauge theory is
strong in the UV, and we will indeed check this prediction later on. 

For large number of D3 branes, the tadpole contribution can be neglected
and an approximate solution is \cite{kt}
\be
\Phi^{(0)} = \Phi_0 \ , \ e^{\xi^{(0)}} = e^{\Phi_0} + \sqrt{2} M \rho \ ,
\ e^{\eta^{(0)}} = 2 \sqrt{\rho} \ . \label{b5}
\ee 
In the near horizon limit, this reduces to the $AdS_5 \times S^5$
background. The AdS/CFT correspondence identifies the energy scale $u$ of
the gauge theory to $u^2 = exp(-\xi / 2)$.

In the next approximation, we can compute the linear fluctuation around
this solution due to the tadpole
\be
\Phi = \Phi^{(0)}+ \Phi^{(1)} \ , \ 
\xi = \xi^{(0)}+ \xi^{(1)} \ , \
\eta = \eta^{(0)}+ \eta^{(1)} \ . \label{b6}
\ee 
In the IR limit ($u << 1$) $\rho \rightarrow \infty$, a 
straightforward computation gives
\ba
&\Phi^{(1)}& = - {3 \alpha \over 64} (\sqrt{2} M) e^{3 \Phi_0 \over
2} \ln \rho \ , \nonumber \\
&\xi^{(1)} (\rho)& = -{\alpha \over 128} e^{3 \Phi_0 \over 2} 
(\sqrt{2} M)^{1 \over 2} + {C \over \rho}  \ , \nonumber \\
&\eta^{(1)} (\rho)& =  -{\alpha \over 128} e^{3 \Phi_0 \over 2} 
(\sqrt{2} M)^{1 \over 2} + {C' \over \rho} 
 \ , \label{b7}
\ea 
where $\Phi_0$,$C$ and $C'$ are integration constants. We did choose some
other integration constants in (\ref{b7}) in order to keep the linear
fluctuations small in the IR.
    
The physical gauge coupling is $1/g_{YM}^2 = M exp(\Phi)$, which is
approximately given, in the IR region, by neglecting for simplicity the
linear fluctuation in $\xi$, by
\be
{1 \over g_{YM}^2} = e^{-\Phi_0} -{3 \alpha \over 16 M} (\sqrt{2} M
e^{\Phi_0})^{1 \over 2} \ln u \ . \label{b8}
\ee
The logarithmic evolution of the gauge couplings, coming from the
dilaton tadpole and noticed previously in
\cite{aa} is remarkable, since is qualitatively similar to the
renormalization group running of the D3 brane gauge theory. Notice,
however, that since $\alpha >0$, the gauge theory is predicted to be
free in the IR and not in the UV, as previously thought. This prediction
can be readily checked by computing the gauge theory one-loop beta function for
the D3 gauge theory using the spectrum displayed in (\ref{a061}) and
in Section 4.1. Notice first of all from (\ref{wz3}) that the $U(1)$
gauge boson in $U(M)$ mixes with the two-form $A_2$ and becomes massive,
leaving an $SU(M)$ unbroken gauge group. Using standard formulae, we
then find the one-loop beta function 
\ba
&b_{SU(M)}& = {11 \over 3} T_G - {2 \over 3} \sum_f T_f - {1 \over 3} 
\sum_s T_s = \nonumber \\
&& {11 \over 3} M - {8 \over 3} ({M-2 \over 2}+{M+2 \over 2}) - {2 \over
3} \times {32 \over 2} - {1 \over 6} \times {6 M} = 
- {32 \over 3} \ , \label{b9}
\ea  
where $T_G$ denotes the Dynkin index for the adjoint representation and
$T_f$ ($T_s$) is the Dynkin index for Weyl fermion (complex scalar) 
representations. The result (\ref{b9})
indeed confirms the sign and the $1/M$ dependence appearing on the gauge 
theory side, found by taking the large $M$ limit with fixed $M
exp(\Phi)$. Moreover,
the connection between the gauge theory one-loop beta function and the
dilaton tadpole is transparent, since because on the gauge theory side the
D3-D3 spectrum is conformal at one-loop and only the D9-D3 spectrum, containing
Weyl fermions in the representation ${\bf (32,M)}$, contributes. 
The numerical factor $32$ is here the rank of the D9 brane gauge group $U(32)$,
which determines also the dilaton tadpole (\ref{a0600}). 
The precise numerical coefficient, however, in (\ref{b8}) is different
from the one appearing in (\ref{b9}). This is due to the fact,
noticed in \cite{kt}, that corrections to (\ref{b8}) are of
the same order as the leading result.  

In the UV region $u >>1$, the gauge coupling becomes strong. Field
eqs. (\ref{b04}) can be analytically solved if we retain only the
tadpole contribution. However, the solution depends on a large number of
integration constants which should be fixed by matching the IR solution
(\ref{b5}), (\ref{b7}) with the UV one. This seems hard to realize, but
it would be interesting to see the fate of the Landau pole on the
gravity side. 
 
\section{Conclusion}

 In this paper we have determined the spectrum of D-branes in the
non-tachyonic 0'B orientifold using
two methods: the first is based on solving the consistency
conditions which are essentially $\Omega^2=1$ on the 
tensor product of the Chan-Paton and oscillator
degrees of freedom and correct transverse channel interpretation. 
The second consists in 
considering compactifications on $T^{2n}/Z_2$
where in order to cancel the tadpoles one has to
introduce D-branes. By moving the branes from the fixed points
and taking the infinite volume limit one gets the corresponding
branes in non-compact space.

 We have found that all the D branes are  characterised by world-volume
$U(M)$ gauge groups with a chiral fermion content. There are
massless Weyl fermions in the symmetric, antisymmetric and fundamental 
representations of $U(M)$, the chiralities 
for the antisymmetric and fundamental representations
being the same. We verified that the world-volume anomalies
associated to the  gauge and gravitational anomalies due to
the chiral fermions are precisely those requested by the
form of the ten-dimensional anomaly and by the Green-Schwarz
mechanism.  This also allowed us to predict new Wess-Zumino
couplings of the D-branes.  

We also found a 9d compactification which in the closed sector
interpolates between IIB and OB. By choosing a Klein projection similar
to the 10d O'B model \cite{augusto}, we eliminate the closed string
tachyon for {\it any} radius. The open sector is tachyon free for all
radii, too, and has a spectrum combining in an interesting way
supersymmetry breaking in the bulk (M-theory breaking \cite{ads2}) and
supersymmetry breaking on the branes (brane supersymmetry breaking
\cite{ads}).

 Another relation between the world-volume  and  
 spacetime theories is the Maldacena conjecture that we
 considered for the D-3 branes. The gauge theory of the latter has 
 a positive beta function which is independent of $N$ at one loop.
 We related the running of the gauge coupling constant to the
 classical variation of the dilaton due to the disk tadpole, with 
 a good qualitative agreement. 

Due to the Ramond-Ramond charges carried and the non-tachyonic nature of
their spectrum, the D branes discussed in this
paper are expected to be stable, even if they are not BPS. 
However, the existence of the NS-NS tadpoles and also the nonvanishing
of the one-loop vacuum energy means that the perturbative vacuum is not
the true ground state. In particular, this means that while for BPS
branes, far away (with the exception of D7 and D9
branes) the background is asymptotic to the
Minkowski space, this is not the case for the type of non-BPS branes
discussed in our paper.
One possibility (in the case of non-compact space) is to take a large number of
branes, in which case the tadpoles and one-loop vacuum energy are
subleading in most of the transverse space. Another one is to find the 
explicit background redefinition, in
the spirit of the Fishler-Susskind mechanism \cite{fs}, as in the
10d example worked recently in \cite{dm}. This second possibility,
discussed briefly in Section 6, together with the quantization around
the new background, clearly deserves further attention and work.

\vskip 16pt
\begin{flushleft}
{\large \bf Acknowledgments}
\end{flushleft}

\noindent We are grateful to C. Angelantonj and A. Sagnotti for useful
discussions and comments. E.D. would like to thank the Aspen Center
for Physics for hospitality and J.M. is grateful to the Center for
Advanced Mathematical Sciences in Beirut, where part of this work was 
carried out.  


\appendix
\section{The Wess-Zumino terms of the D7 and D5 branes}

The anomaly polynomial of  the $D7$-brane
is given by
\be
I_{10}=\hat A(\Sigma)[
e^{{{N}\over{2}}}Tr_A(e^{iG})-e^{-{{N}\over{2}}}
Tr_S(e^{iG})+tr(e^{iF})tr(e^{iG})] \ . \label{and7} 
\ee
In this Appendix we explain some steps that lead 
to the factorised form
(\ref{an7}) and determine explicitly the terms dependent on the
normal curvature $N$ in the $Y_i$.
First, we note that $\hat A(\Sigma)$ can be written as
\be
\hat A(\Sigma)=\hat A(R)\left(\hat A(N)\right)^{-1} \ ,
\ee
where $R$ is the pull-back of spacetime curvature
on the brane. Since the normal bundle is one-dimensional, the
roof genus $\hat A(N)$ is given by
\be
\hat A={{N/2}\over{sinh(N/2)}} \ .
\ee
Using this relation, as well as
\be
Tr_A^Se^{iG}={{1}\over{2}}\left[ (tr e^{iG})^2\pm
tr(e^{2iG})\right] \ , \label{trr}
\ee
the anomaly (\ref{and7}) can be put in the form
\be
I_{10}=\hat A(R)\left[tr e^{iG}
{{sinh(N/2)}\over{N/2}}\left(tr e^{iF} + {{N}\over{2}}
{{sinh(N/2)}\over{N/2}}\right)-tr e^{2iG}
{{sinh(N)}\over{N}}\right] \ . \label{annn}
\ee
If we define $J= tr e^{iG} (2/N) \ {sinh(N/2)}  = \sum J_{2i}$, where
$J_{2i}$ is the $2i$ form part, we have
\be
tr e^{2iG} {{sinh(N)} \over{N}}=\sum 2^iJ_{2i} \ .
\ee
Inserting this into (\ref{annn}), we get
\be
I_{10}=\hat A(R)\left[J\left(tr e^{iF}+ {{N}\over{2}}J \right)
-\sum 2^iJ_{2i}\right] \ .
\ee
Now we use the expansion of the roof genus
\be
\hat A=1-{{p_1}\over{24}}+{{7p_1^2-4p_2}\over{5760}}+\dots \ , \label{roof}
\ee
where we defined the first two Pontryagin classes
\be
p_1 (R) = - { 1 \over 2} tr R^2 \ , \ p_2 (R) = {1 \over 8} (tr R^2)^2
- {1 \over 4} tr R^4 \ . \label{pontryagin}
\ee
By using also the expression of the polynomials $X_i$ displayed in 
(\ref{ans1})
\ba
X_8&=&{{1}\over{4!}} tr F^4-{{1}\over{48}}p_2+{{1}\over{64}}p_1^2
+ {{1}\over{96}}tr F^2p_1 \ , \nonumber\\
X_{10}&=&{{i}\over{5!}}tr F^5-{{1}\over{2880}} itrFp_2-
{{1}\over{3840}} itrF p_1^2 \ , 
\ea
we can put the anomaly in the form
\ba
I_{10}&=&MX_{10}+J_2X_8+(J_4-M{{p_1}\over{24}})X_6
+(J_6-J_2{{p_1}\over{48}})X_4\nonumber\\
&+&(J_8-{{M}\over{2880}}p_2-{{M}\over{3840}}p_1^2)X_2\nonumber\\
&+&N(MJ_8+J_2J_6+{{1}\over{2}}J_4^2-{{M}\over{24}}J_4p_1
-{{1}\over{48}}J_2^2p_1+M^2{{7p_1^2-4p_2}\over{5760}}) \ .
\label{fact7}
\ea
From the eq. (\ref{fact7}), we deduce
\ba
Y_2&=&J_2=itr G, \ Y_4=J_4-M{{p_1}\over{24}}=
-{{1}\over{2}}tr G^2+ {{M} \over {24}}N^2+{{M}\over{48}}tr
R^2 \ , \nonumber\\
Y_6&=&J_6-J_2{{p_1}\over{48}}=-{{i}\over{6}}tr G^3+{{i}\over{24}}
tr G N^2+{{i}\over{96}}trG tr R^2 \ , \nonumber\\
Y_8&=& J_8 - {M \over 2880}p_2 - {M \over 48 \times 80} p_1^2 = \nonumber \\
&& {{1}\over{24}}tr G^4-{{1}\over{48}}tr G^2
N^2+{{M}\over{1920}}N^4+{{M}\over{11520}}tr R^4
-{{M}\over{9216}}(tr R^2)^2 \ .
\ea
The term which multiplies $N$ in the last line 
in (\ref{fact7}) 
can be put in the form
\be
MY_8+Y_2Y_6+{{1}\over{2}}Y_4^2 \ ,
\ee
which completes the derivation of (\ref{an7}).

The anomaly of the 5-brane
reads
\be
I_6=\hat A(\Sigma) [ch_+(N)Tr_A(e^{iG})-ch_-(N)
Tr_S(e^{iG})+tr(e^{iF})tr(e^{iG})] \ . \label{and5} 
\ee
Let $\lambda_1$ and $\lambda_2$ be the 
Chern roots of the curvature in the fundamental 
representation of $SO(4)$, then
\begin{equation}
ch(S_{\pm})=e^{({\lambda_1\pm\lambda_2 \over 2})}+
 e^{-({\lambda_1\pm\lambda_2 \over 2})} \ .
\end{equation}
The first Pontryagin class is given by 
$p_1=\lambda_1^2+\lambda_2^2$, the Euler 
class is given by $\chi=\lambda_1\lambda_2$ 
and the second Pontryagin class by $p_2=\lambda_1^2\lambda_2^2$.
Therefore we get the Chern characters of the spin bundle 
\begin{equation}
ch(S_{\pm})=2+{{(p_1\pm 2\chi)}\over{4}}+{{p_1^2+4p_2\pm 
4p_1\chi}\over{192}} \ .
\end{equation}
The roof genus of the normal bundle is now given by
\be
\hat A(N)^{-1}={{sinh(\lambda_1/2) \ sinh(\lambda_2/2)}\over{
\lambda_1\lambda_2/4}} \ ,
\ee
so that using again (\ref{trr}), the anomaly polynomial reads
\be
I_{8}=\hat A(R)\left[K\left(tr e^{iF}+ {{\lambda_1\lambda_2}
\over{2}}K \right)
-\sum 2^{i+1}K_{2i}\right] \ , \label{an5i}
\ee
where
\be
K={{sinh(\lambda_1/2)sinh(\lambda_2/2)}\over{
\lambda_1\lambda_2/4}} \ tr e^{iG} \ .
\ee
Equation (\ref{an5i}) can be cast in the form
\ba
I_8&=&MX_8+K_2X_6+\left(K_4-{{M}\over{48}}p_1(R)\right)X_4
+K_6X_2\nonumber\\
&+&\lambda_1\lambda_2\left(
-{{M^2}\over{48}}p_1(R)+MK_4+{{1}\over{2}}
K_2^2\right).\label{lan}
\ea
We easily deduce the $Y$ polynomials, given by
\ba
Y_2&\!\!=\!\!&K_2=itr G \ , \ Y_4=K_4-{{M}\over{48}}p_1(R) \!=\!
\!-\!{{1}\over{2}}trG^2+{{M}\over{24}}p_1(N)\!+\!{{M}\over{96}} tr R^2 \ ,
\nonumber\\
Y_6&=&K_6=-{{i}\over{6}}tr G^3+{{i}\over{24}}trG \ p_1(N).
\ea
The last line in (\ref{lan}) can be cast in the form
\be
\chi(N) (MY_4+{{1}\over{2}}Y_2^2),
\ee
which completes the derivation of (\ref{an5}).




\begin{thebibliography}{99}

\bibitem{polchinski} J. Polchinski, String Theory, Cambridge
Univ. Press, 1998.

\bibitem{maldacena} J.M. Maldacena, {\it Adv. Theor. Math. Phys.}{\bf 2}
(1998) 231; S.S. Gubser, I.R. Klebanov and A.M. Polyakov,
\PLB{428}{98}{105}; E. Witten, {\it Adv. Theor. Math. Phys.}{\bf 2}
(1998) 253. 

\bibitem{polyakov} A.M. Polyakov, hep-th/9809057.

\bibitem{kt} I.R. Klebanov and A.A. Tseytlin, \NPB{546}{99}{155};
\JHEP{9903}{99}{015}; \NPB{547}{99}{143}; J.A. Minahan, \JHEP{9901}{99}{020};
I.R. Klebanov, N.A. Nekrasov and S. Shatashvili, hep-th/9909109; 
A. Armoni, B. Kol, \JHEP{9907}{99}{011}.

\bibitem{bfl2} R. Blumenhagen, A. Font and D. L\"ust, \NPB{560}{99}{66};
A. Armoni, E. Fuchs and J. Sonnenschein, \JHEP{9906}{99}{027};
R. Blumenhagen, C. Kounnas, D. L\"ust  {\it J. High Energy Phys.} {\bf
0001} (2000) 036. 

\bibitem{augusto} A. Sagnotti, hep-th/9702093; hep-th/9509080.

\bibitem{carlo} C. Angelantonj, \PLB{444}{98}{309}.

\bibitem{bfl} R. Blumenhagen, A. Font and D. L\"ust, \NPB{558}{99}{159}. 

\bibitem{aa} C. Angelantonj and A. Armoni, {\it Nucl. Phys. } {\bf B578} (2000) 239;
{\it Phys.Lett.} {\bf B482} (2000) 329; M. Bianchi and J. F. Morales, 
{\it J. High Energy Phys.} {\bf 0008} (2000) 035.   

\bibitem{bg} O. Bergman and M.R. Gaberdiel, \JHEP{9907}{99}{022}.

\bibitem{cargese} A. Sagnotti, in: Cargese '87, Non-Perturbative Quantum
Field Theory, eds. G. Mack et al. (Pergamon Press, Oxford, 1988) p.
521;  P. Horava, \NPB{327}{89}{461}.

\bibitem{ps} G. Pradisi and A. Sagnotti, \PLB{216}{89}{59}; M. Bianchi,
G. Pradisi and A. Sagnotti, \NPB{376}{92}{365}.

\bibitem{bs} M. Bianchi and A. Sagnotti, \PLB{247}{90}{517};
\NPB{361}{91}{519}.

\bibitem{gp} E. Gimon and J. Polchinski, \PRD{54}{96}{1667}.

\bibitem{sw} N. Seiberg and E. Witten, \NPB{276}{86}{272}.

\bibitem{gsw} M. Green, J. Schwarz and E. Witten, Superstring theory,
Cambridge Univ. Press, 1987.

\bibitem{agg} L. Alvarez-Gaum\'e and P. Ginsparg, \AP{161}{85}{423}. 

\bibitem{pst} P. Pasti, D. Sorokin and M. Tonin, \PRD{55}{97}{6292}. 

\bibitem{hw} C.M. Hull, E. Witten, \PLB{149}{84}{117}

\bibitem{ddp} J.A. Dixon, M.J. Duff, J.C. Plefka,
\PRL{69}{92}{3009}.

\bibitem{lt} K. Lechner, M. Tonin, \NPB{475}{96}{545}

\bibitem{mourad} J. Mourad, \NPB{512}{98}{199}. 

\bibitem{lm}K. Lechner, P.A. Marchetti, hep-th/0007076.

\bibitem{pw}J. Polchinski, E. Witten, \NPB{460}{96}{525}.

\bibitem{Li} M. Li, \NPB{460}{96}{351}; M.R. Douglas, hep-th/9512077;
M.B. Green, J.A. Harvey and G. Moore, {\it Class. Quantum Grav.}
{\bf 14} (1997) 47; J.F. Morales, C.A. Scrucca and M. Serone, 
\NPB{552}{99}{291}; B. Stefanski, \NPB{548}{99}{275}.  

\bibitem{bk} R. Blumenhagen and A. Kumar, \PLB{464}{99}{46}. 

\bibitem{sugimoto} S. Sugimoto, \PTP{102}{99}{685}.

\bibitem{ads} I. Antoniadis, E. Dudas and A. Sagnotti,
\PLB{464}{99}{38}; G. Aldazabal and A.M. Uranga, hep-th/9908072;
G. Aldazabal, L.E. Ib{\'a}n\~ez and F. Quevedo, hep-th/9909172;
C. Angelantonj, I. Antoniadis, G. D'Appollonio, E. Dudas and
A. Sagnotti, {\it Nucl. Phys.} {\bf B572} (2000) 36.
 
\bibitem{fs} W. Fischler and L. Susskind, \PLB{171}{86}{383}; 
\PLB{173}{86}{262}. 

\bibitem{dm} E. Dudas and J. Mourad, {\it Phys. Lett.} {\bf B486}
(2000) 172.

\bibitem{ads2}  I. Antoniadis, E. Dudas and A. Sagnotti,
\NPB{544}{99}{469}.

\bibitem{bw} R. Bott and L. Wu, Differential forms in algebraic
geometry, Springer-Verlag, 1983.

\end{thebibliography}
\end{document}